\documentclass[]{aastex63}

\usepackage{natbib}
\usepackage{booktabs}
\usepackage{rotating, graphicx}
\usepackage{verbatim}
\usepackage{longtable}
\usepackage{comment}
\def\lax {\ifmmode{_<\atop^{\sim}}\else{${_<\atop^{\sim}}$}\fi}
\def\gax {\ifmmode{_>\atop^{\sim}}\else{${_>\atop^{\sim}}$}\fi}
\def\gtorder{\mathrel{\raise.3ex\hbox{$>$}\mkern-14mu
             \lower0.6ex\hbox{$\sim$}}}

\def\NH{N_\mathrm{H}}

\def\me{m_\mathrm{e}}
\def\kte{kT_\mathrm{e}}
\def\kts{kT_\mathrm{s}}
\def\ktbb{kT_\mathrm{bb}}

\def\me{m_\mathrm{e}}
\def\Kalpha{K_\mathrm{\alpha}}

\def\chiq{$\chi^2$}
\def\chiqr{$\chi^{2}_\mathrm{red}$}
\def\s1{s$^{-1}$}
\def\cm2{cm$^{-2}$}
\def\ster1{ster$^{-1}$}

\def\xspec{{\sc xspec}}

\def\xmm{{\it XMM-Newton}\/}

\def\INT{{\it INTEGRAL}} 
\def\SUZ{{\it SUZAKU}}
\def\Ch{{\it Chandra}}
\def\eg{e.g., }
\def\compTT{{\sc compTT}\/}

\shorttitle{Comptonization spectra of IPs}
\shortauthors{Maiolino et al.}

\begin{document}

\title{Comptonization as an origin of the continuum 
in Intermediate Polars}

\author{T. Maiolino}
\affiliation{School of Physics and Technology, Wuhan University, Wuhan
430072, China}
\affiliation{WHU-NAOC Joint Center for Astronomy, Wuhan University, Wuhan 430072, China}
\affiliation{Dipartimento di Fisica, Universit\`a di Ferrara, via Saragat 1, 44122 Ferrara, Italy}

\author{L. Titarchuk}
\affiliation{Astro Space Center, Lebedev Physical Institute, Russian
Academy of Sciences,  Profsouznay ul. 84/32, Moscow 117997, Russia}

\author{W. Wang}
\affiliation{School of Physics and Technology, Wuhan University, Wuhan
430072, China}
\affiliation{WHU-NAOC Joint Center for Astronomy,
Wuhan University, Wuhan 430072, China}

\author{F. Frontera}
\affiliation{Dipartimento di Fisica, Universit\`a di Ferrara, via Saragat
1, 44122 Ferrara, Italy}
\affiliation{INAF/OAS Bologna, via Gobetti 101, 40129 Bologna, Italy}
\affiliation{ICRANET Piazzale d. Repubblica 10-12, 65122 Pescara (PE), Italy}

\author{M. Orlandini}
\affiliation{INAF/OAS Bologna, via Gobetti 101, 40129 Bologna, Italy}

\correspondingauthor{Wang Wei}
\email{wangwei2017@whu.edu.cn}
   
\begin{abstract}
In this paper we test if the $\sim$ 0.3 -- 15 keV {\xmm} EPIC pn spectral continuum of IPs can be described by the thermal Comptonization \textsc{compTT} model. 
We used publicly observations of 12 IPs (AE~Aqr, EX Hya, V1025 Cen, V2731 Oph, RX J2133.7+5107, 
PQ Gem, NY Lup, and  V2400 Oph, IGR~J00234+6141, IGR~J17195-4100, V1223 Sgr, and XY~Ari). We find that our modeling is capable to fit well the average spectral continuum 
of these sources. In this framework, UV/soft X-ray seed photons (with $<\kts>$ of 0.096 $\pm$ 0.013 keV) coming presumably from the star surface are scattered off 
by electrons present in an optically thick plasma (with $<\kte>$ of 3.05 $\pm$ 0.16 keV and optical depth $<\tau>$ of 9.5 $\pm$ 0.6 for plane geometry) located nearby (on top) to the more central seed photon  emission regions. A soft blackbody (\textsc{bbody}) component is observed in 5 out of the 13 observations analysed, with a mean temperature $<kT_{bb}>$ of $0.095 \pm 0.004$ keV.  
We observed that the spectra of IPs show in general two photon indices $\Gamma$, which are driven by the source luminosity and optical depth. Low luminosity IPs show $<\Gamma>$ of $1.83 \pm 0.19$, whereas high luminosity IPs show lower $<\Gamma>$ of $1.34 \pm 0.02$. 
Moreover, the good spectral fits of PQ Gem and V2400 Oph indicate that the polar subclass of CVs may be successfully described by the thermal Comptonization as well. 
  
\end{abstract} 

\keywords{X-rays: binaries --- cataclysmic variables --- 
accretion, radiation mechanisms: thermal -- scattering}

\section{Introduction}
\label{sec:int}

Intermediate polars (IPs) are asynchronous rotating ($P_{spin} < P_{orb}$) magnetic Cataclysmic Variables (mCVs).  In these systems, the magnetic field ($B \sim 10^{6-7}$ G) of the white dwarf (WD) is strong enough  to influence the accretion flow, disrupting the accretion disk. 
 The accretion flow is expected to attach to the magnetic field lines of the WD at the magnetosphere radius -- being directed towards the magnetic pole caps of the star. 
It has been suggested that the accretion flow forms the ``accretion curtain'' region above the magnetic poles. In this region the accretion flow is geometrically thin, arc-shaped and tall \citep{Kuulkers2006}. 
Some IPs are expected to accrete like in the polar subclass of mCVs --  that is, directly through a stream (disk-less or stream-fed IPs). The accreting stream 
can overflow the disk (disk-overflow accretion) or completely replace it \citep[see, e.g., ][and references therein]{Warner1995,Kuulkers2006}.

It is currently accepted that hard X-rays are formed  due to the shock produced by the interaction of  
the material in the accreting column with the WD. The post-shock region cools 
emitting hard X-rays and its emission has been described by thermal bremsstrahlung or thermal bremsstrahlung based models. 
In the frame of this standard scenario, it has been suggested that the plasma temperature of this region  is typically in the $\sim$ 5 -- 60~keV range \citep{Hellier2001,Warner1995}.

The existing analyses of IPs (like in nonmagnetic CVs (nmCVs)) are  
inhomogeneous: different spectral models have been used to describe their X-ray spectra.
  For example: \citet{Mukai2003} demonstrated that a 
 \textit{Chandra} spectrum of the IP EX~Hya is well fitted by a cooling-flow model, as well as the spectra of the DNe U~Gem, SS~Cyg and the old nova V603~Aql. In contrast, the 
\textit{Chandra} spectra of the IPs V1223~Srg, AO~Psc and GK~Per are inconsistent with such a model; \citet{Landi2009} presented a spectral analysis of 22 CVs observed with  {\INT}\ \textit{IBIS} (20 -- 100~keV energy range). Their analysis indicated that the best-fit model is a 
thermal bremsstrahlung with an average temperature $<kT> \sim 22$~keV. The authors obtained similar results (temperatures in the $\sim$ 16 -- 35 keV range) when 
combining not simultaneous ($\sim$ 0.3 -- 100 keV) \textit{Swift/XRT} and IBIS 
data of 11 sources. \citet{Xu2016} analysed 16 IPs using \SUZ\ data. They modeled their spectra with the single temperature optically thin thermal plasma \textsc{apec} model (with metalicity set to zero) and obtained temperatures all above $\sim$15~keV.

Summing up, the spectral models used to fit IPs spectra are basically the same as 
those used to fit nmCVs spectra: 
 the continuum is described either by bremsstrahlung 
or by thermal optically thin plasma models, like the \textsc{apec} 
code\footnote{Calculated using the \textsc{atomdb} code, more information can be 
found at \url{http://atomdb.org/}; and their variations \textsc{vapec} and \textsc{vvapec}.} or \textsc{mekal} code (which simultaneously  model the emission lines) and their variations. If required by the data, either more than one optically thin plasma temperature  or cooling flow model is used to describe the data. It is important to notice that all these models have bremsstrahlung emission 
as their basis describing the continuum emission.

In mCVs a soft X-ray component has been observed in the broad-band X-ray spectrum. 
This component was discovered by \textit{ROSAT} and is  
typically described by a blackbody emission with temperatures ranging from a few up to $\sim$ 100 eV 
\citep{deMartino2004,deMartino2006a,deMartino2006b,Mason1992,Burwitz1996}. 
In IPs this soft X-ray emission is believed to originate just like in polars, that is, around the accreting poles by the reprocessing of the hard emission on the WD surface  \citep{deMartino2004,deMartino2006a,deMartino2006b}. The blackbody fluxes indicate fractional areas of only 
$\sim10^{-5}$ of the WD surface for the soft X-ray emission region 
\citep{Harbel1995}. Interestingly, not all  
IPs show this blackbody component. \citet{Evans2007}, performing a systematic spectral analysis of \xmm\  
data, stated that this soft blackbody is likely a common component in the X-ray spectra of IPs, suggesting  
that the lack of the observation of this component can in general be explained by geometrical effects. Namely, depending on the system inclination and the magnetic colatitude (see their Figure 5), the polar region and hence the blackbody component could be hidden by the accretion curtain.

In this standard spectral modelling, the X-ray emission can appear  highly absorbed within the accretion flow (with $N_H\sim10^{23}$ cm$^{-2}$). Therefore, fits including an extra cold medium totally covering the source or partially ionized covering absorbers (which varies with the orbital phase) are commonly found in the literature \citep[see, \eg][]{Norton1989,Ishida1994,Landi2009}. A reflection component (from the WD surface) is also expected and used in the spectral modelling of IPs \citep[see, \eg][]{Cropper1998,Beardmore2000}.

Fluorescent Fe $K_\alpha$ lines have been observed in most of the IPs 
\citep[see, \eg][]{Norton1991,Ishida1991,Hellier1998,Xu2016}.  
The moderate spectral resolution of instruments on board \textit{Chandra}, \xmm\  and
\textit{Suzaku} shows that usually up to three emission lines are present 
in the $\sim$ 6.4 -- 7.0 keV range of the spectra. These lines are  usually modeled by simple Gaussian components and their centroid 
energies usually 
correspond to the cold 
(neutral) fluorescent Fe $K_\alpha$ line at 6.4 keV, and the He-   
and H-like Fe lines at 6.7 and 7.0 keV, respectively. It is important to notice, however, that more complex   
line profiles have been reported in these sources \citep[see, \eg][]{Ishida1992,Titarchuk2009}. In this standard framework, these lines are believed to be produced by irradiation (or reflection) of hard X-rays on material 
present on the star surface or somewhere in the accretion flow near the compact object:  in the 
pre-shock material -- this interpretation is based on the observation of Doppler-shifted red-wing (redshifted) line in \eg GK~Per -- and/or in the lower velocity base of the accretion columns -- this interpretation is based on the absence of Doppler shifts in the H- and He-like components \citep{Hellier2004}. 

As one can clearly see, the Comptonization  process is not the standard one  which has been used to describe the X-ray continuum of IPs. We tested if the thermal Comptonization could successfully  
describe the X-ray continuum of these sources  based on the following points:   
(1) the standard spectral modelling of IPs is found to be similar to nmCVs  
(2) the continua of nmCVs can be successfully described by thermal Comptonization 
\citep{Maiolino2020} (Paper I hereafter) 
(3) the \xmm\ Epic pn spectra of the IP GK~Per is successfully fitted by the thermal \compTT\ model in XSPEC  \citep{Titarchuk2009}, and 
(4) IPs partially shares geometric/structural similarities with both nmCVs and low mass X-ray binaries (LMXBs). 

Accordingly with the catalog of IPs and IP candidates provided by NASA 
(\url{http://asd.gsfc.nasa.gov/Koji.Mukai/iphome/catalog/alpha.html}), 
eleven out of the 12 IPs present in our sample (AE~Aqr, EX Hya, V1025 Cen, V2731 Oph, RX J2133.7+5107, 
PQ Gem, NY Lup, V2400 Oph, IGR~J00234+6141, V1223 Sgr, and XY~Ari)  are classified as ironclad IPs, while 1 (IGR~J17195-4100) is a confirmed IP.

This paper is organized as follows: in section  
\ref{sec:datared} we shortly present the \xmm\ data reduction, 
in section \ref{sec:specan} 
we show the results of  the spectral analysis, in section \ref{sec:disc} we discuss our results, demonstrate  some examples of the spectral modelling in the standard framework for some sources in our sample, and present the new thermal  Comptonization modelling  for  IPs.
Finally, in section \ref{sec:conc} we summarize our main results and conclusions.

\section{Data Reduction}
\label{sec:datared}

The \xmm\ Epic-pn data reduction was performed through the Science Analysis Software SAS version 14.0.0.
Table \ref{tab:obs} shows the log of the 13  observations present in our sample. All observations 
are taken in imaging mode, which allows the extraction of spectra and light curves in the 
0.3 to 15~keV energy range. 
All spectra and light curves were extracted strictly following  
the \textit{Users Guide to the \xmm\ Science Analysis System} 
\cite{XMMuserguide}, and the \textit{\xmm\ ABC Guide} \citep{XMMABCGUIDE}.

Standard filters were applied on the EVL through the \texttt{evselect} 
task:  taking only good events from the PN camera (using \#XMMEA\_EP), 
using single and doubles events (i.e. using pattern $\leq4$) in the energy 
range selected, and omitting parts of the detector area like border pixels 
(and columns with higher offset) for which the patter type and the total 
energy is known with significantly lower precision (using \texttt{FLAG==0}).   

We used the SAS task \textit{epatplot} to check for pile-up. We found that only the observations of AE Aqr 
and V2400 Oph were affected by pile-up. In these two cases, in order to mitigate 
this effect, an inner circle of radius equal to 150 physical unity was excised from the source 
extraction region during data reprocessing. 

In the extraction of light curves, the background subtraction was done accordingly through the SAS 
task \texttt{epiclccor}, which performs corrections at once for various effects affecting the detection 
efficiency. 

We investigated spectral variability in each observation of our sample by computing hardness ratios 
(HRs) with light curves produced in 3 energy ranges: 0.3 -- 5.0~keV,  5.0 -- 8.0~keV, and 8.0 -- 15.0~keV. 
We observed significant variation (pulses) only in the 5.0 -- 8.0/0.3 -- 5.0 HR of EX Hya. In this case we have not observed spectral differences (except for differences in the normalization) when fitting its  spectrum with filter in time -- that is, not considering the data taken during the time of the peaks in the 5.0 -- 8.0/0.3 -- 5.0 HR. This result was expected since the pulses are produced by minima in the 0.3 -- 5.0 keV light curve. 
 
Therefore, we have considered the total EPIC-pn exposure time in all final spectral analyses. 
The photon redistribution matrix (rmf) and the ancillary (arf) files were created through the \texttt{rmfgen} 
and \texttt{arfgen} task, respectively. All spectra were rebinned in order to have at least 25 counts for each background-subtracted channel.

\begin{table}
\caption{Log of the \textit{XMM-Newton} Epic-pn Observations of IPs in our sample}
\label{tab:obs}
\centering
\footnotesize
\begin{tabular}{lllllll}
\toprule
Source &Obs.~ID &Data Mode &Start Time (UTC) &End Time (UTC) & Exp (ks)\\
\midrule
AE Aqr &0111180201 &Imaging &2001-11-07~~23:45:53 &2001-11-08~~03:38:17 &12.3 \\
EX Hya &0111020101 &Imaging &2000-07-01~~08:07:39 &2000-07-01~~20:05:10  &30.2\\
V1025 Cen &0673140501 &Imaging &2012-01-01~~22:50:42 &2012-01-02~~02:48:30 &13.3\\
V2731 Oph &0302100201 &Imaging &2005-08-29~~07:05:40 &2005-08-29~~10:13:31 &10.5 \\
RX J2133.7+5107 &0302100101&Imaging &2005-05-29~~13:32:59 &2005-05-29~~17:19:24 &12.7 \\
PQ Gem &0109510301 &Imaging &2002-10-07~~23:13:51 &2002-10-08~~09:07:21 &24.9 \\
NY Lup &0105460301 &Imaging &2000-09-07~~10:43:41 &2000-09-07~~16:10:31 &17.3 \\
NY Lup &0105460501 &Imaging &2001-08-24~~16:35:06 &2001-08-24~~20:46:58 &13.5 \\  
V2400 Oph &0105460601 &Imaging &2001-08-30~~15:36:48 &2001-08-30~~19:56:46 &13.7 \\
IGR J00234+6141 &0501230201 &Imaging & 2007-07-10~~05:58:25 &2007-07-10~~12:54:50 &21.7 \\  
IGR J17195-4100 &0601270201 &Imaging & 2009-09-03~~06:57:59 &2009-09-03~~15:51:53 &27.6 \\  
V1223 Sgr &0145050101 &Imaging & 2003-04-13~~12:40:18 & 2003-04-13~~23:22:08 &27.0 \\  
XY Ari    &0501370101 &Imaging & 2008-02-14~~18:46:45 & 2008-02-15~~04:16:08 &25.6 \\         
\bottomrule 
\end{tabular}
\end{table}

\section{Spectral Analysis}
\label{sec:specan}

The spectral analysis was performed with the \xspec\ astrophysical spectral package \citep{Arnaud1996} version 12.8.2. 
The 0.3 to 15 keV spectral continua were modeled with 
a soft blackbody component (when required by the data) plus a thermal 
Comptonization component, 
modified by a photo-electric absorption (\textsc{tbabs} model in XSPEC)
due to the presence of absorber material (hydrogen 
column $\NH$) in the line of sight of the source.

The blackbody component was modeled by the \textsc{bbody} 
model in XSPEC, and the Comptonization component was modeled by the thermal \compTT\   
model in XSPEC \citep{Titarchuk1994,Hua1995,Titarchuk1995}, considering a cylindrical (plane) geometry.
In this model, soft seed photons of temperature $\kts$ are scattered off by  electrons present in a hot 
plasma (Compton cloud) with temperature $\kte$ and Thomson optical depth $\tau$.  

When residuals were evident in the $\sim$ 6.4 to 7.0 keV energy range, up to three Gaussian components (named in our analysis as Gaussian$_1$, Gaussian$_2$ and Gaussian$_3$) were added to the total model to account for the neutral, He- and H-like Fe lines usually observed in this energy range. When additional spectral features were observed (mainly in the E $\lesssim$ 2.5 keV band) extra Gaussian components (named in our analysis as Gaussian$_{A,B,C,...,L.}$) were added to the total model.  
 
The best-fitting parameters of the continuum and fit quality are shown in Table \ref{tab:cont}. The Gaussian components of the Fe complex are shown in Table \ref{tab:Felines}, whereas the other lines are shown in Table \ref{tab:lines}.

\begin{sidewaystable}
\centering
\caption{Spectral analysis using the thermal \textsc{compTT} model to describe the $\sim$ 0.3 to 15.0 keV continuum of the IPs. The \textsc{Gaussian} components are shown in Table \ref{tab:Felines} and \ref{tab:lines}.}
\label{tab:cont}
\normalsize
\begin{tabular}{lllllllllllll}  
Component &TBABS & &\multicolumn{2}{c}{\textsc{bbody}} & &\multicolumn{4}{c}{\textsc{compTT}} & \chiqr/d.o.f & $\Delta$$E^{f}$ (keV)\\ 
        \cline{4-5} \cline{7-10} 
 Parameters &$\NH$ (10$^{22}$$^{a}$) & &kT$_{bb}$ (keV)   &norm$^{g}$ & &kT$_{s}$ (keV) &kT$_{e}$ (keV) &$\tau$ &norm$^{g}$ & & & \\
 Source & &  & & & & & & & & & &\\
 \toprule
AE~Aqr          &$5.37\times10^{-2}$$^{b}$ & & & & &$0.060_{-0.011}^{+0.002}$ &$2.0_{-c}^{+0.5}$ &$5.9_{-0.8}^{+0.1}$ &$7.3_{-1.3}^{+0.2}$ &1.19/549 &$0.3-15$\\
\vspace{1.5mm}
EX Hya          &$0.28_{-0.05}^{+0.04}$ & &$0.089_{-0.003}^{+0.003}$ &$22_{-4}^{+6}$ & &$0.10^{b}$ &$3.05_{-0.14}^{+0.17}$ &$6.35_{-0.26}^{+0.26}$ &$26.7_{-0.8}^{+0.8}$ &1.18/1726 &$1.0-15$\\
\vspace{1.5mm}
V1025 Cen       &0.01$_{-c}^{+0.03}$ & & & & &$0.181_{-0.029}^{+0.014}$ &$3.1_{-0.3}^{+0.6}$ &$7.1_{-0.7}^{+0.5}$ &$2.5_{-0.4}^{+0.4}$ &1.15/398 &$0.3-15$\\
\vspace{1.5mm}
V2731~Oph       &$0.156^{b}$ & & & & &$0.018_{-0.003}^{+0.004}$ &$3.20_{-0.18}^{+0.22}$ &$12.6_{-0.4}^{+0.4}$ &$2.51_{-0.11}^{+0.11}$ &1.06/696 &$1.0-15$\\
\vspace{1.5mm}
RX~J2133.7+5107 &$0.134_{-0.013}^{+0.013}$ & &$0.105_{-0.002}^{+0.003}$ &$0.175_{-0.023}^{+0.027}$ & &$0.17_{-0.08}^{+0.04}$ &$3.30_{-0.19}^{+0.23}$ &$10.4_{-0.4}^{+0.4}$ &$1.95_{-0.15}^{+1.00}$ &1.12/1047 &$0.3-15$\\
\vspace{1.5mm}
PQ~Gem          &$0.033^{b}$ & & & & &$0.13_{-c}^{+0.04}$ &$2.72_{-0.08}^{+0.09}$ &$11.69_{-0.24}^{+0.23}$ &$2.40_{-0.19}^{+1.05}$ &0.96/1461 &$0.5-15$\\
\vspace{1.5mm}
V2400~Oph       &$0.004_{-c}^{+0.007}$ & & & & &$0.108_{-0.011}^{+0.010}$ &$2.37_{-0.08}^{+0.08}$ &$11.8_{-0.3}^{+0.3}$ &$7.2_{-0.3}^{+0.3}$ &1.04/1235 &$0.3-15$\\
\vspace{1.5mm}
IGR J00234+6141 &0.167$_{-0.011}^{+0.011}$ & & & & &0.134$_{-0.014}^{+0.013}$ &$2.84_{-0.21}^{+0.26}$ &9.8$_{-0.5}^{+0.6}$ &$0.87_{-0.07}^{+0.06}$ &0.92/725 &$0.3-15$\\
\vspace{1.5mm}
IGR J17195-4100 &0.091$_{-0.006}^{+0.006}$ & & & & &$0.100_{-0.014}^{+0.013}$ &2.73$_{-0.07}^{+0.07}$ &10.64$_{-0.20}^{+0.20}$ &$4.88_{-0.21}^{+0.27}$ &1.13/1703 &$0.3-15$\\
\vspace{1.5mm}
XY Ari          &$4.29_{-0.17}^{+0.23}$ & & & & &0.1$^{b}$ &4.29$_{-0.28}^{+0.41}$ &$8.0_{-0.6}^{+0.5}$ &$2.13_{-0.05}^{+0.05}$ &1.12/837 &$0.3-15$\\
\vspace{1.5mm}
V1223 Sgr       &$0.51_{-0.05}^{+0.05}$ & &$0.10_{-0.01}^{+0.01}$ &$2.2_{-0.6}^{+0.1}$ & &0.109$^{b}$ &$2.97_{-0.09}^{+0.10}$ &$10.0_{-0.3}^{+0.3}$ &$10.8_{-0.2}^{+0.2}$ &1.06/1870 &$1.0-15$\\
\vspace{1.5mm}
NY~Lup$^{d}$    &$0.100_{-0.013}^{+0.010}$ & &$0.095_{-0.004}^{+0.003}$ &$0.1186_{-0.0013}^{+0.0012}$ & &$0.068_{-0.006}^{+0.007}$ &$3.13_{-0.13}^{+0.15}$ &$10.28_{-0.28}^{+0.28}$ &$3.46_{-0.03}^{+0.03}$ &1.02/1373 &$0.3-15$\\
\vspace{1.5mm}
NY~Lup$^{e}$    &0.218$^{b}$ & &$0.084_{-0.002}^{+0.002}$ &$0.95_{-0.05}^{+0.06}$ & &$0.021_{-0.004}^{+0.004}$ &$3.9_{-0.3}^{+0.3}$ &$9.1_{-0.3}^{+0.3}$ &$5.7_{-0.4}^{+0.3}$ &1.04/1281 &$0.5-15$\\
\bottomrule
\end{tabular}
\begin{itemize}
\item[ ] Uncertainties at 90\% confidence level 
\item[a] atoms.cm$^{-2}$
\item[b] Parameter frozen
\item[c] Parameter pegged at hard limit
\item[d] Obs. ID 0105460301
\item[e] Obs. ID 0105460501
\item[f] Spectral energy range 
\item[g] $\times 10^{-03}$
\end{itemize} 
\end{sidewaystable}

\begin{sidewaystable}
\centering
\caption{\textsc{Gaussian} components present in the $\sim 6.4 - 7.0$ keV energy range of the Fe emission line complex.}
\label{tab:Felines}
\footnotesize
\begin{tabular}{lllllllllllllll}
 & \multicolumn{4}{c}{Gaussian$_1$} & & \multicolumn{4}{c}{Gaussian$_2$} & & \multicolumn{4}{c}{Gaussian$_3$}\\
 \cline{2-5} \cline{7-10} \cline{12-15}
 &E$_1$ &$\sigma_1$ & norm$_{1}^{e}$ & EW$_1$ & &E$_2$ &$\sigma_2$ &norm$_2^{e}$ &EW$_2$ & &E$_3$ &$\sigma_3$ &norm$_3^{e}$ &EW$_3$\\
Source          & & & & & & & & & & & & & &\\
\toprule
AE~Aqr          & & & & & &$6.690_{-0.027}^{+0.026}$ &$0.05^{+0.04}_{-c}$ &$1.5_{-0.3}^{+0.3}$ &$821_{-60}^{+431}$ & & & & &\\ 
\vspace{1.5mm}
EX Hya          &$6.39_{-0.02}^{+0.02}$ &$0.02^{b}$ &$3.1_{-0.6}^{+0.6}$ &$24_{-5}^{+6}$ & &$6.659_{-0.003}^{+0.003}$ &$0.04^{b}$ &$33.7_{-1.0}^{+1.0}$ &$350_{-11}^{+13}$ & &$6.96_{-0.01}^{+0.01}$ &$0.02^b$ &$12.1_{-0.7}^{+0.7}$ &$108_{-6}^{+8}$\\
\vspace{1.5mm}
V1025 Cen       & & & & & &6.78$_{-0.04}^{+0.04}$ &0.17$_{-0.04}^{+0.04}$ &4.5$_{-0.9}^{+0.9}$ &$324_{-67}^{+83}$ & & & & &\\
\vspace{1.5mm}
V2731~Oph       &$6.42_{-0.10}^{+0.08}$ &$0.39_{-0.11}^{+0.12}$ &$10.5_{-2.2}^{+2.5}$ &$472_{-92}^{+102}$ & & & & & & & & & &\\
\vspace{1.5mm}
RX~J2133.7+5107 &$6.431_{-0.022}^{+0.023}$ &$0.11_{-0.03}^{+0.03}$ &$7_{-1}^{+1}$ &$268_{-31}^{+238}$ &  & & & & & &$6.89_{-0.04}^{+0.04}$ &$0.07^{b}$ &$2.2_{-0.7}^{+0.7}$ &$93_{-26}^{+77}$ \\ 
\vspace{1.5mm}
PQ~Gem          &$6.38_{-0.02}^{+0.04}$ &$0.10_{-0.04}^{+0.06}$ &$4_{-1}^{+1}$ &$154_{-32}^{+74}$ & &$6.66_{-0.03}^{+0.04}$ &$0.02^{b}$ &$1.9_{-1.0}^{+0.6}$ &$63_{-23}^{+38}$ & &$6.94_{-0.03}^{+0.03}$ &$0.0c2^{b}$ &$1.5_{-0.4}^{+0.4}$ &$58_{-16}^{+31}$\\
\vspace{1.5mm}
V2400~Oph       &$6.39_{-0.02}^{+0.04}$ &$0.07_{-0.06}^{+0.06}$ &$10_{-2}^{+3}$ &$150_{-35}^{+93}$ & &$6.7_{-0.04}^{+0.06}$ &$0.07^{b}$ &$7_{-2}^{+2}$ &$95_{-30}^{+36}$ & &$6.96_{-0.03}^{+0.04}$ &$0.07^{b}$ &$5.6_{-1.5}^{+1.4}$ &$98_{-29}^{+32}$\\
\vspace{1.5mm}
IGR J00234+6141 & & & & & &6.70$_{-0.05}^{+0.05}$ &0.23$_{-0.04}^{+0.05}$ &2.7$_{-0.5}^{+0.4}$ &$386_{-9}^{+13}$ & & & & &\\
\vspace{1.5mm}
IGR J17195-4100 &$6.39_{-0.02}^{+0.02}$ &$0.07^{b}$ &$5.2_{-0.7}^{+0.3}$ &$54_{-12}^{+6}$ & &$6.70_{-0.02}^{+0.03}$ &$0.07^{b}$ &$2.9_{-0.7}^{+0.3}$ &$119_{-19}^{+20}$ & &$6.97_{-0.02}^{+0.03}$ &$0.07^{b}$ &$2.9_{-0.6}^{+0.6}$ &$81_{-15}^{+17}$\\
\vspace{1.5mm}
XY Ari          &$6.43_{-0.05}^{+0.05}$ &$0.07^{b}$ &$1.6_{-0.6}^{+0.7}$ &$63_{-27}^{+47}$ & &6.68$_{-0.04}^{+0.04}$ &$0.07^{b}$ &$2.4_{-0.7}^{+0.7}$ &$103_{-26}^{+69}$ & &$7.00_{-0.04}^{+0.04}$ &$0.07^{b}$ &$1.4_{-0.5}^{+0.6}$ &$66_{-23}^{+45}$\\
\vspace{1.5mm}
V1223 Sgr       &$6.40_{-0.01}^{+0.01}$ &$0.02^{b}$ &$9.4_{-0.8}^{+0.8}$ &$88_{-8}^{+9}$ & &$6.68_{-0.01}^{+0.01}$ &$0.02^{b}$ &$8.2_{-0.8}^{+0.8}$ &$73_{-7}^{+8}$ & &$6.96_{-0.02}^{+0.02}$ &$0.05^{b}$ &$6.9_{-0.8}^{+0.9}$ &$71_{-9}^{+9}$\\
\vspace{1.5mm}
NY~Lup$^{c}$    &$6.38_{-0.02}^{+0.02}$ &$0.07$$^{b}$ &$5.1_{-0.7}^{+0.7}$ &$145_{-25}^{+29}$ & &$6.67_{-0.03}^{+0.03}$ &$0.07$$^{b}$ &$4.2_{-0.7}^{+0.7}$ &$111_{-21}^{+27}$ & &$6.94_{-0.03}^{+0.03}$ &$0.07$$^{b}$ &$3.6_{-0.7}^{+0.7}$ &$112_{-25}^{+29}$\\
\vspace{1.5mm}
NY~Lup$^{d}$    &$6.37_{-0.02}^{+0.02}$ &$0.07^{b}$ &$5.5_{-0.9}^{+0.9}$ &$132_{-23}^{+21}$ & &$6.68_{-0.02}^{+0.02}$ &$0.07$$^{b}$ &$6.4_{-0.9}^{+0.9}$ &$139_{-23}^{+25}$ & &$6.97_{-0.02}^{+0.02}$ &$0.04^{b}$ &$5.5_{-0.8}^{+0.8}$ &$120_{-20}^{+23}$\\
\bottomrule 
\end{tabular}
\begin{itemize}
\item[ ] Uncertainties at 90\% confidence level 
\item[a] Parameter frozen
\item[b] Parameter pegged at hard limit
\item[c] Obs. ID 0105460301
\item[d] Obs. ID 0105460501
\item[e] $\times 10^{-5}$
\end{itemize} 
\end{sidewaystable}

\begin{table}[h]
\centering
\caption{Other lines observed in the spectrum the IPs in our sample.}
\label{tab:lines}
\scriptsize
\begin{tabular}{lllllllllll} 
Component& Parameter &Unit &AE~Aqr &EX Hya &V1025 Cen &PQ~Gem &V2400~Oph &IGR~J17195-4100 & NY~Lup$^{a}$ & NY~Lup$^{b}$\\ 
\toprule
Gaussian$_A$ &E$_{A}$    &keV & & & &0.52$_{-0.02}^{+0.02}$ & &$0.57_{-0.01}^{+0.01}$ & &\\     
\vspace{1.5mm}                                
             &$\sigma_A$ &keV & & & &$0.10_{-0.02}^{+0.02}$ & &$0.05_{-0.02}^{+0.02}$ & &\\     
\vspace{1.5mm}      
            &EW$_A$      &eV & & & &$492_{-75}^{+59}$ & &$54_{-12}^{+6}$ & &\\
Gaussian$_B$ & E$_{B}$   &keV &$0.83_{-0.01}^{+0.01}$ & &0.844$_{-0.012}^{+0.012}$  &$0.89_{-0.02}^{+0.02}$ &0.808$_{-0.014}^{+0.016}$ & & &\\     
\vspace{1.5mm}                                
             &$\sigma_B$ &keV &$0.17_{-0.01}^{+0.01}$ & &$0.02^{c}$ &$0.08_{-0.04}^{+0.06}$ &0.059$_{-0.023}^{+0.023}$ & & &\\     
\vspace{1.5mm}      
             &EW$_B$     &eV &$608_{-127}^{+4}$ & &$24_{-6}^{+7}$  &$57_{-10}^{+50}$ &-34$^{+12}_{-13}$ & & &\\
Gaussian$_C$ &E$_{C}$    &keV & &$1.107_{-0.007}^{+0.007}$ &1.015$_{-0.018}^{+0.014}$ & & & &$0.98_{-0.02}^{+0.02}$ &$0.99_{-0.04}^{+0.03}$\\
\vspace{1.5mm} 
             &$\sigma_C$ &keV & &$0.02^{c}$ &0.077$_{-0.015}^{+0.018}$ & & & &$0.10_{-0.02}^{+0.02}$ &$0.15_{-0.02}^{+0.03}$\\
\vspace{1.5mm} 
            &EW$_C$      &eV  & &$15_{-2}^{+3}$ &$63_{-12}^{+13}$   & & & &$59_{-13}^{+13}$ &$151_{-27}^{+29}$\\
Gaussian$_D$ &E$_{D}$    &keV & &$1.344_{-0.008}^{+0.008}$ & & & & & &\\
\vspace{1.5mm} 
             &$\sigma_D$ &keV & &$0.02^{c}$ & & & & & &\\
\vspace{1.5mm} 
            &EW$_D$      &eV  & &$14_{-2}^{+2}$ & & & & & &\\
Gaussian$_E$ &E$_{E}$    &keV & &1.473$_{-0.005}^{+0.005}$ & & & & & &\\
\vspace{1.5mm} 
             &$\sigma_E$ &keV & &$0.02^{c}$ & & & & & &\\
\vspace{1.5mm} 
            &EW$_E$      &eV  & &$24_{-2}^{+3}$ & & & & & &\\
Gaussian$_F$ &E$_{F}$    &keV & &$1.848_{-0.008}^{+0.010}$ & & & & & &\\
\vspace{1.5mm} 
             &$\sigma_F$ &keV & &$0.02^{c}$ & & & & & &\\
\vspace{1.5mm} 
            &EW$_F$      &eV  & &$20_{-3}^{+3}$ & & & & & &\\
Gaussian$_G$ &E$_{G}$    &keV & &$1.995_{-0.004}^{+0.006}$ &1.98$_{-0.04}^{+0.04}$ & & & & &\\
\vspace{1.5mm} 
             &$\sigma_G$ &keV & &$0.005^{c}$ &$0.075^{c}$ & & & & &\\
\vspace{1.5mm} 
            &EW$_G$      &eV  & &$28_{-3}^{+2}$ &$34_{-11}^{+11}$ & & & & &\\ 
Gaussian$_H$ &E$_{H}$    &keV & &$2.444_{-0.012}^{+0.011}$ &2.46$_{-0.03}^{+0.04}$ & & & & &\\
\vspace{1.5mm} 
             &$\sigma_H$ &keV & &$0.02^{c}$ &$0.02^{c}$ & & & & &\\
\vspace{1.5mm} 
            &EW$_H$      &eV  & &$19_{-4}^{+4}$ &$38_{-11}^{+12}$  & & & & &\\ 
Gaussian$_I$ &E$_{I}$    &keV & &$2.61_{-0.01}^{+0.01}$ & & & & & &\\
\vspace{1.5mm} 
             &$\sigma_I$ &keV & &$0.02^{c}$ & & & & & &\\
\vspace{1.5mm} 
            &EW$_I$      &eV  & &$24_{-4}^{+3}$ & & & & & &\\ 
Gaussian$_J$ &E$_{J}$    &keV & &$3.918_{-0.024}^{+0.022}$\ & & & & & &\\
\vspace{1.5mm} 
             &$\sigma_J$ &keV & &$0.024^{c}$ & & & & & &\\
\vspace{1.5mm} 
            &EW$_J$      &eV  & &$18_{-4}^{+18}$ & & & & & &\\
Gaussian$_K$ &E$_{K}$    &keV & &$7.82_{-0.02}^{+0.02}$ & & & & & &\\
\vspace{1.5mm} 
             &$\sigma_K$ &keV & &$0.06^{c}$ & & & & & &\\
\vspace{1.5mm} 
            &EW$_K$      &eV  & &$88_{-12}^{+14}$ & & & & & &\\ 
Gaussian$_L$ &E$_{L}$    &keV & &$8.21_{-0.03}^{+0.02}$ & & & & & &\\
\vspace{1.5mm} 
             &$\sigma_L$ &keV & &$0.02^{c}$ & & & & & &\\
\vspace{1.5mm} 
            &EW$_L$      &eV  & &$46_{-12}^{+13}$ & & & & & &\\
\bottomrule 
\end{tabular}
\begin{itemize}
\item[ ] Uncertainties at 90\% confidence level 
\item[a] Obs. ID 0105460301
\item[b] Obs. ID 0105460501
\item[c] Parameter frozen
\end{itemize} 
\end{table}

In all fits, the hydrogen column density $\NH$ was first 
fixed to the value given by the HI4PI survey \citep{HI4PI016}. However, in nine observations (in EX Hya, V1025 Cen, RX~J2133.7+5107, V2400~Oph, IGR~J00234+6141, IGR~J17195-4100, XY~Ari, V1223~Sgr, and NY~Lup Obs. ID 0105460301) a satisfactory fit  
was attained only when this parameter was set to vary freely. 

We noticed that the presence of complex residuals in the soft $\sim$ 0.3 -- 1.0 keV band in EX Hya, V2731~Oph and V1223~Sgr worsened the fit quality. These residuals are likely caused by the contribution of a forest of emission  lines in this band. In V1223~Sgr, for example, lines likely from O VII at 0.574 keV and N VII at $\sim$ 0.5 keV are evident. We observed that the modelling of these features (with Gaussian components) does not strongly affect the best-fitting parameters of the continuum (\compTT\ component), and hence we did not consider this soft energy range in our final fit. 

The presence of the soft \textsc{bbody}  component is necessary in our total spectral modelling of EX Hya, RX~J2133.7+5107, V1223~Sgr and NY~Lup. In EX Hya, RX~J2133.7+5107 and NY~Lup  spectra a fit with reduced \chiq (\chiqr) $<$ 2.0 
is attained only if this component is present. For the other observations we checked the presence of this extra component by means of the F-test defined in terms of the ratio between the normalized \chiq \citep[e.g, ][and references therein]{Maiolino2020}.
In V1223~Sgr the probability of chance improvement (PCI) when adding the extra component is marginal, equal to $1.7\times10^{-3}$. However, when the fit is performed in the total 0.3 -- 15 keV energy band the presence of the soft \textsc{bbody} component is more evident, with PCI of $9\times10^{-5}$  or even lower ($\sim 10^{-19}$) depending on the modelling of the strong and complex residual excesses present in the soft $\sim$ 0.5 -- 1.2 keV energy band. Therefore we decided to keep the \textsc{bbody} component in our final total modelling.

Interestingly, in the analysis of PQ~Gem, a \textsc{bbody} component of $\kts$ of 88$^{+4}_{-3}$ eV is apparently present at first, with PCI of $2.2\times10^{-91}$. However, in this case strong residuals remain in the soft energy band (E $<$ 1 keV), and when Gaussian components are included to take these residuals into account the new PCI for the \textsc{bbody} is equal to 42\%. Therefore, in this case we did not consider the soft \textsc{bbody} component in our final total modelling.

\begin{figure}
\centering
\includegraphics[width=8.5cm]{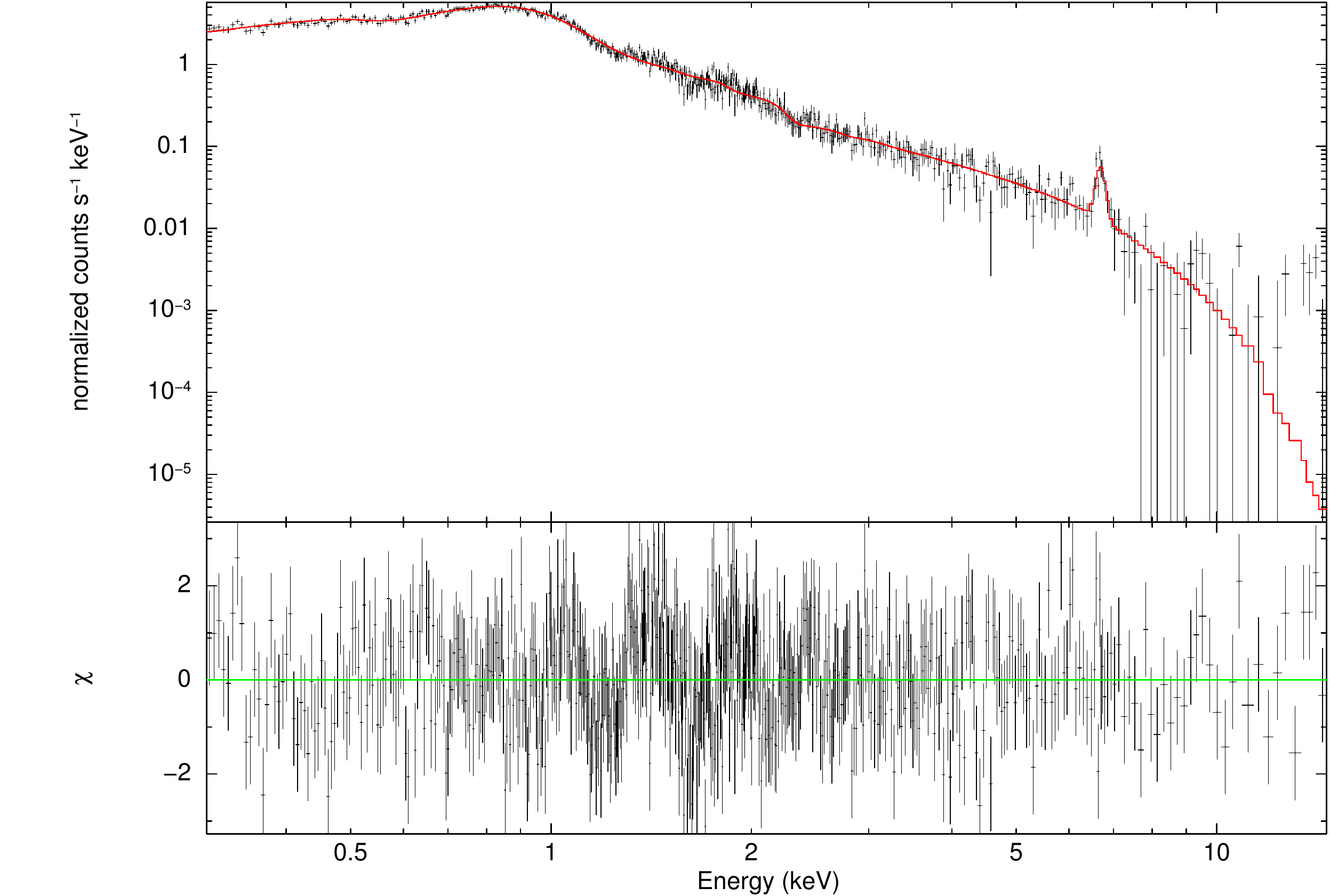}
\includegraphics[width=8.5cm]{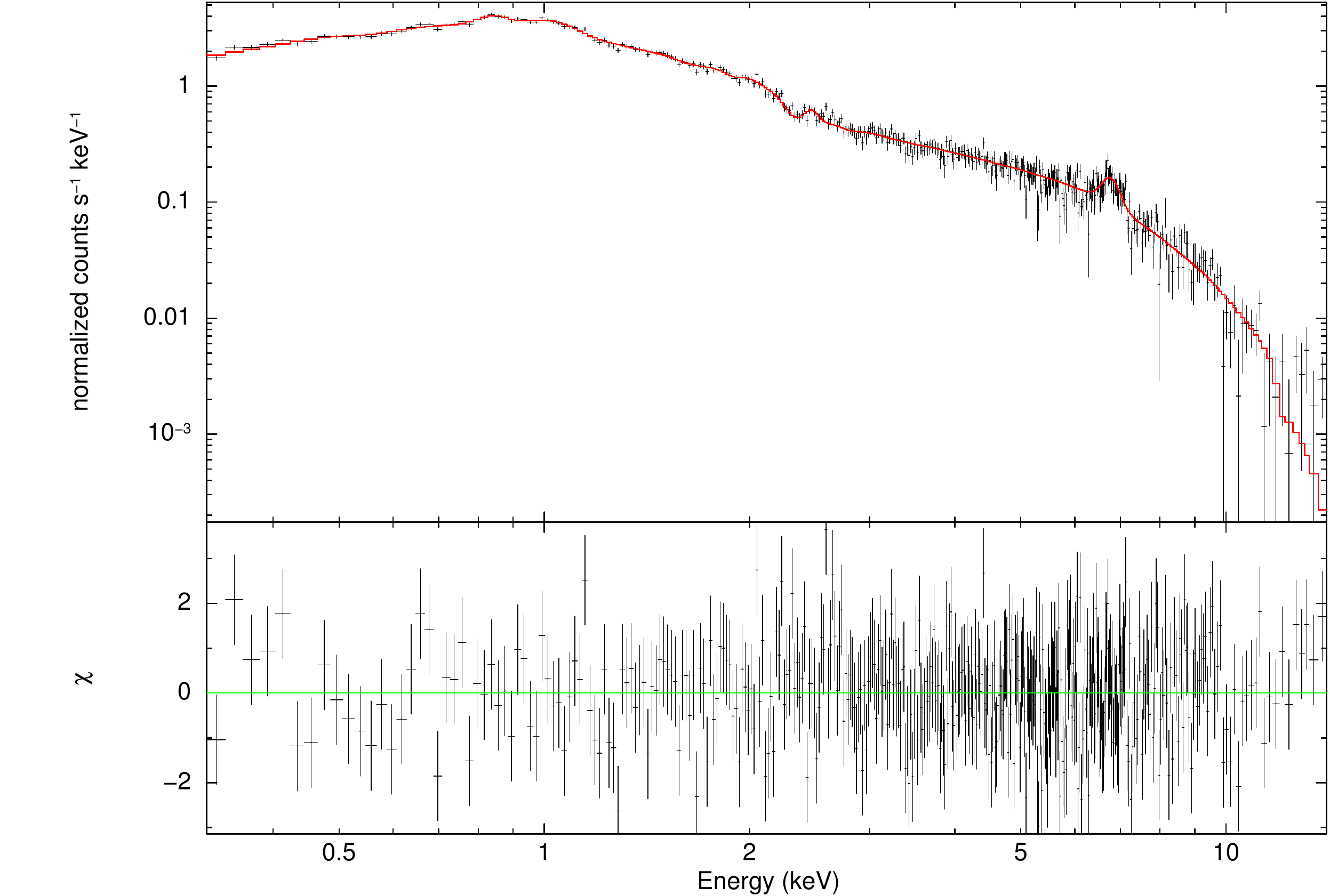}
\includegraphics[width=8.5cm]{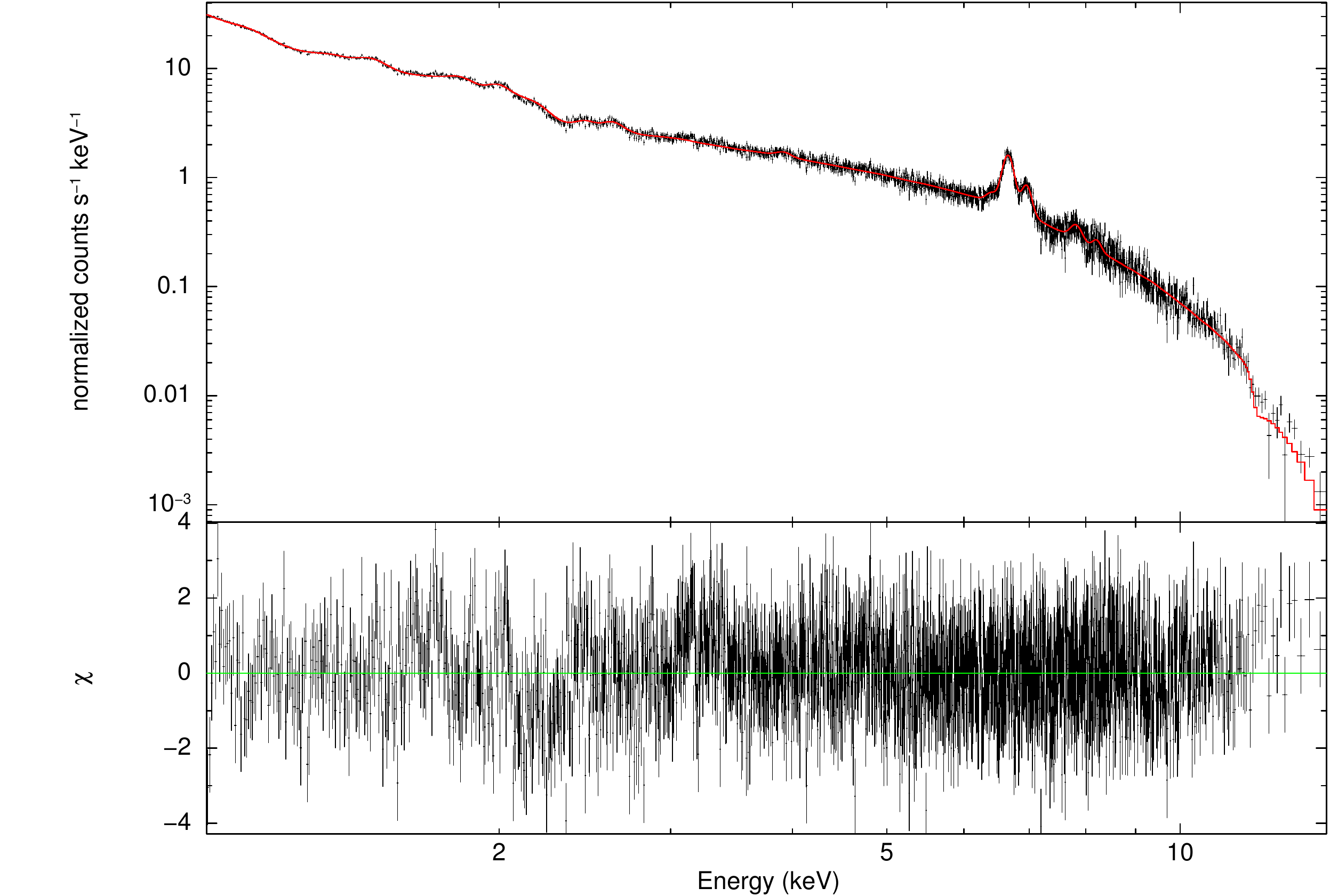}
\includegraphics[width=8.5cm]{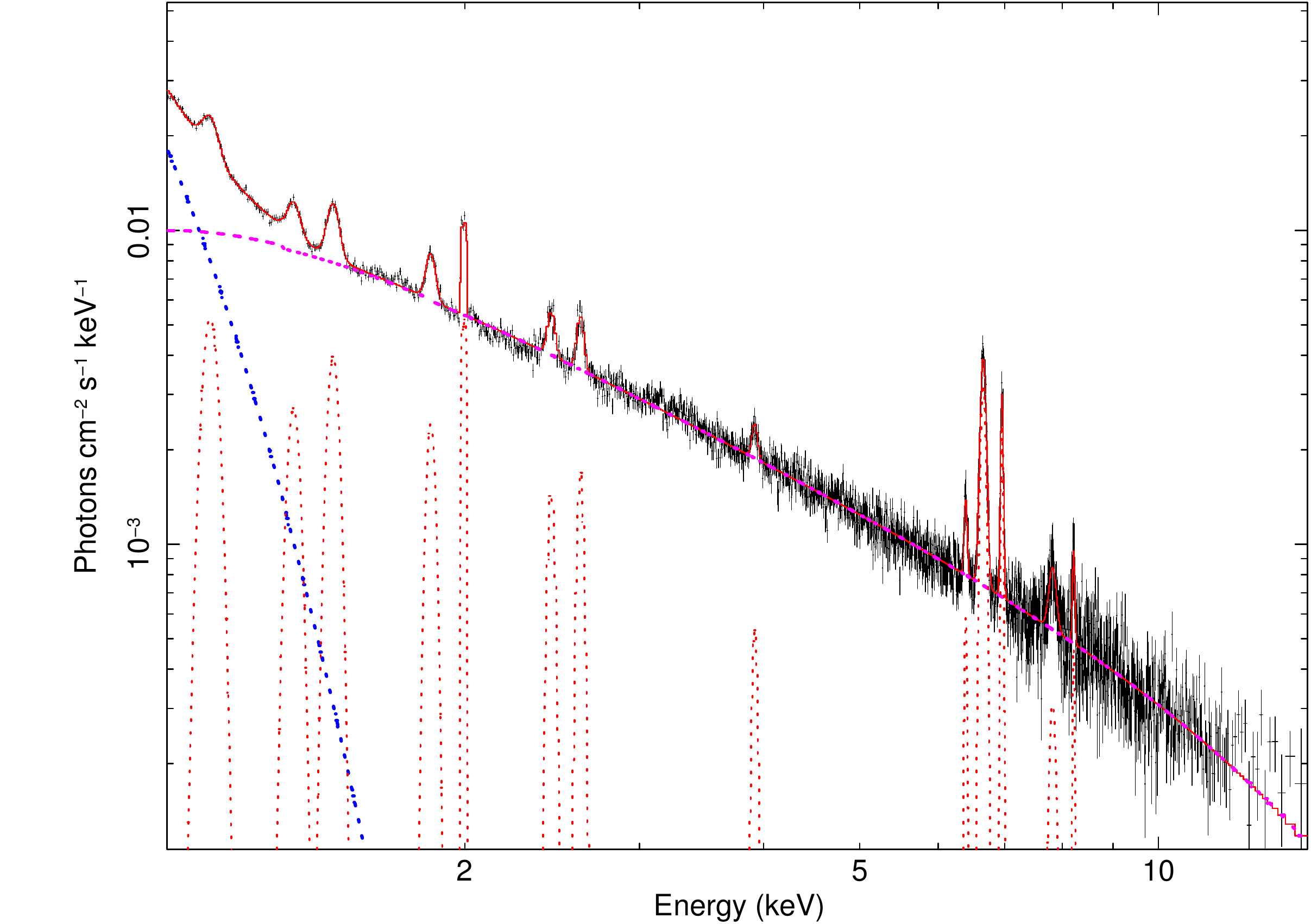}
\caption 
{\xmm\ EPIC pn spectra (\textit{in black}) and the best fit total model (\textit{solid red line}). 
   \textit{Left top panel}: spectrum  
   of AE~Aqr in the 0.3 -- 15.0 keV energy range, and the best fit total model \textsc{tbabs*[compTT+(2)gaussian components]}. 
   \textit{Right top panel}: spectrum of V1025 in the {0.3 -- 15.0} keV energy range, and the best fit total model \textsc{tbabs*[compTT+(5)gaussian components]}. 
   \textit{Left lower panel}: spectrum
   of EX Hya in the 1.0 -- 15.0 keV energy range, and the best fit total model \textsc{tbabs*[compTT+(13)gaussian components]}.  
   \textit{Right lower panel}: unfolded spectrum
   of EX Hya in the 1.0 -- 15.0 keV energy range; the total model in red, the \textsc{bbody} component in blue, the \compTT\ component in magenta, and the \textsc{Gaussian} components in dashed red lines. See Table \ref{tab:cont}, \ref{tab:Felines}, and \ref{tab:lines}. 
}
\label{fig:spec1XMM}
\end{figure}

\begin{figure}
\centering
\includegraphics[width=8.5cm]{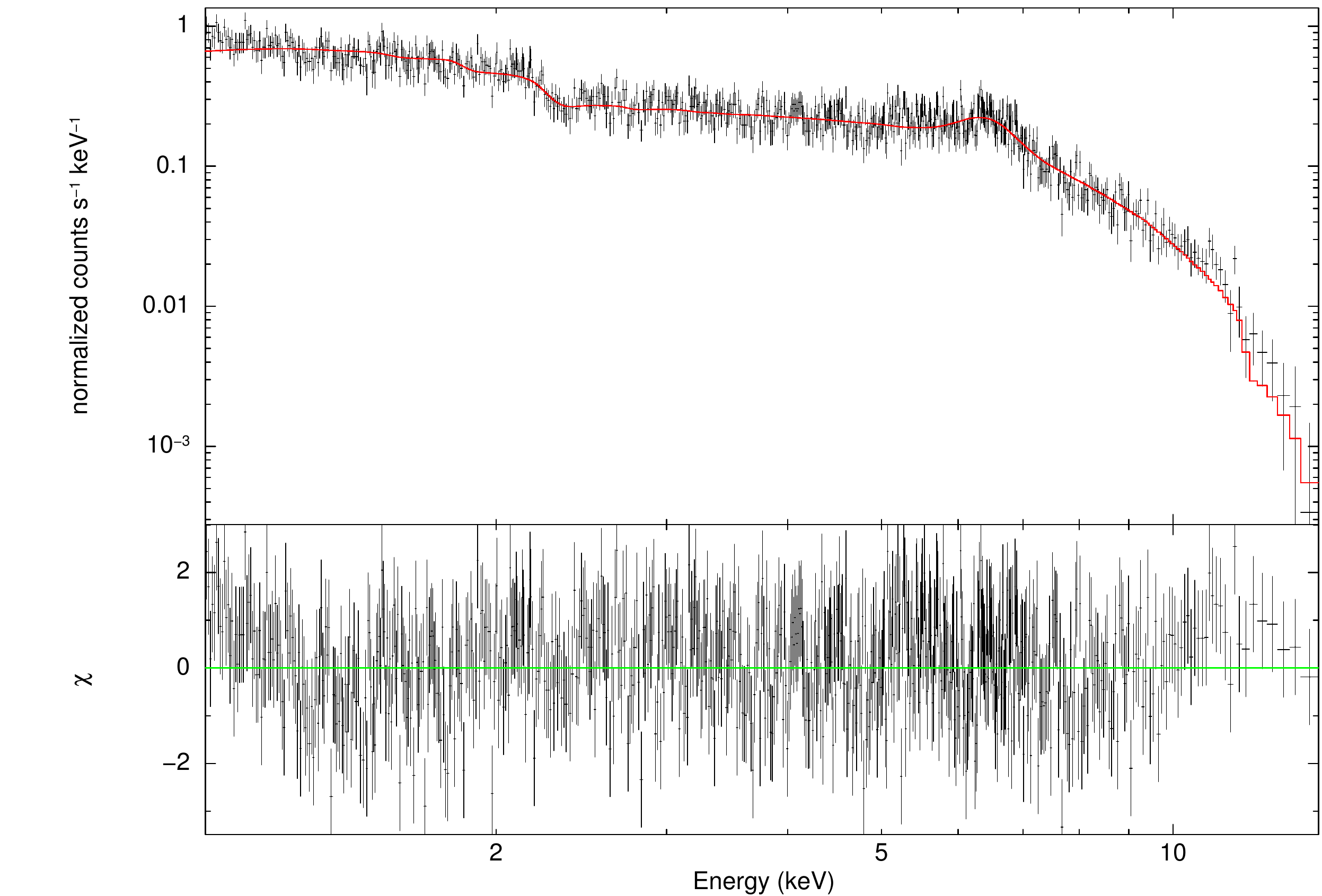}
\includegraphics[width=8.5cm]{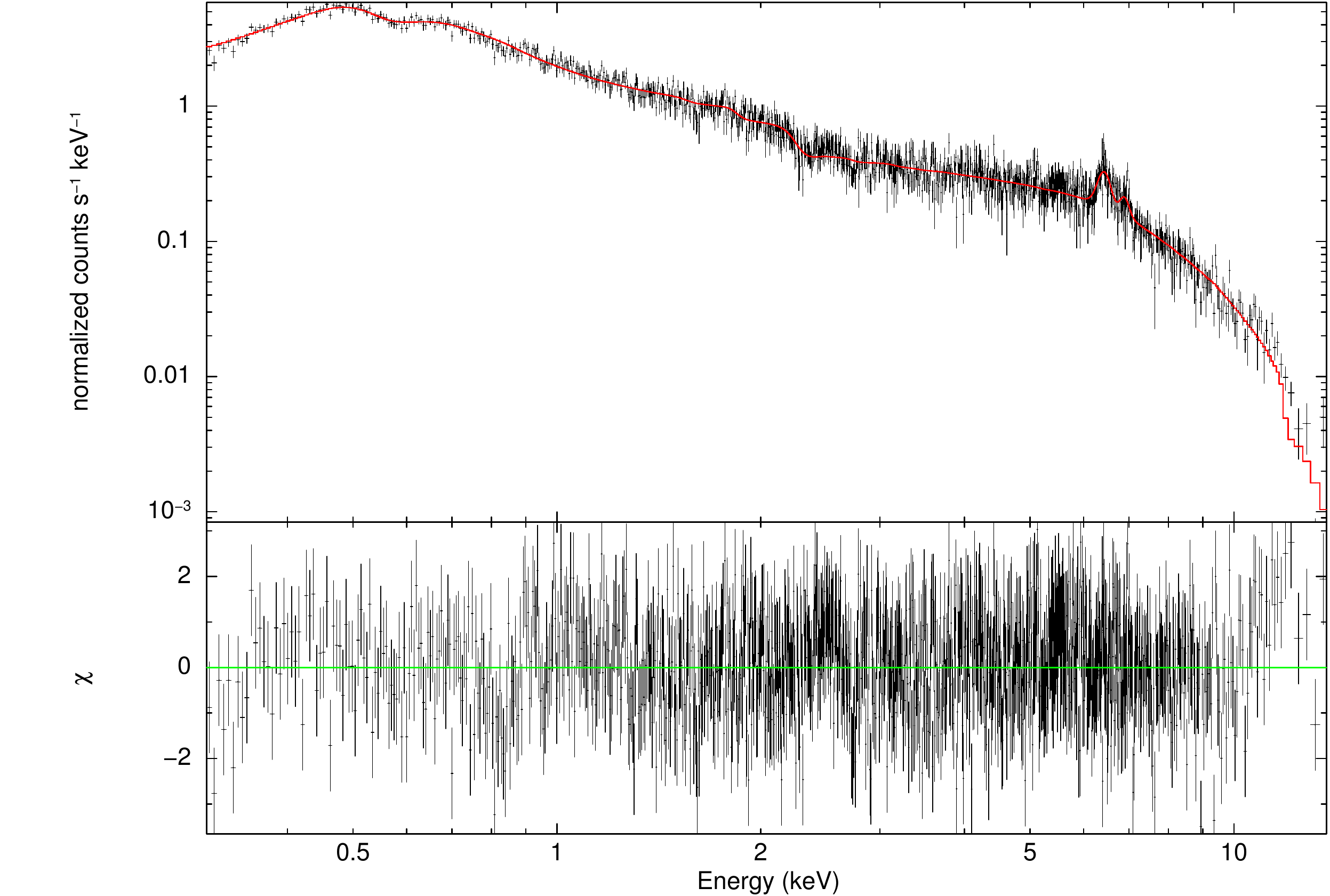}
\includegraphics[width=8.5cm]{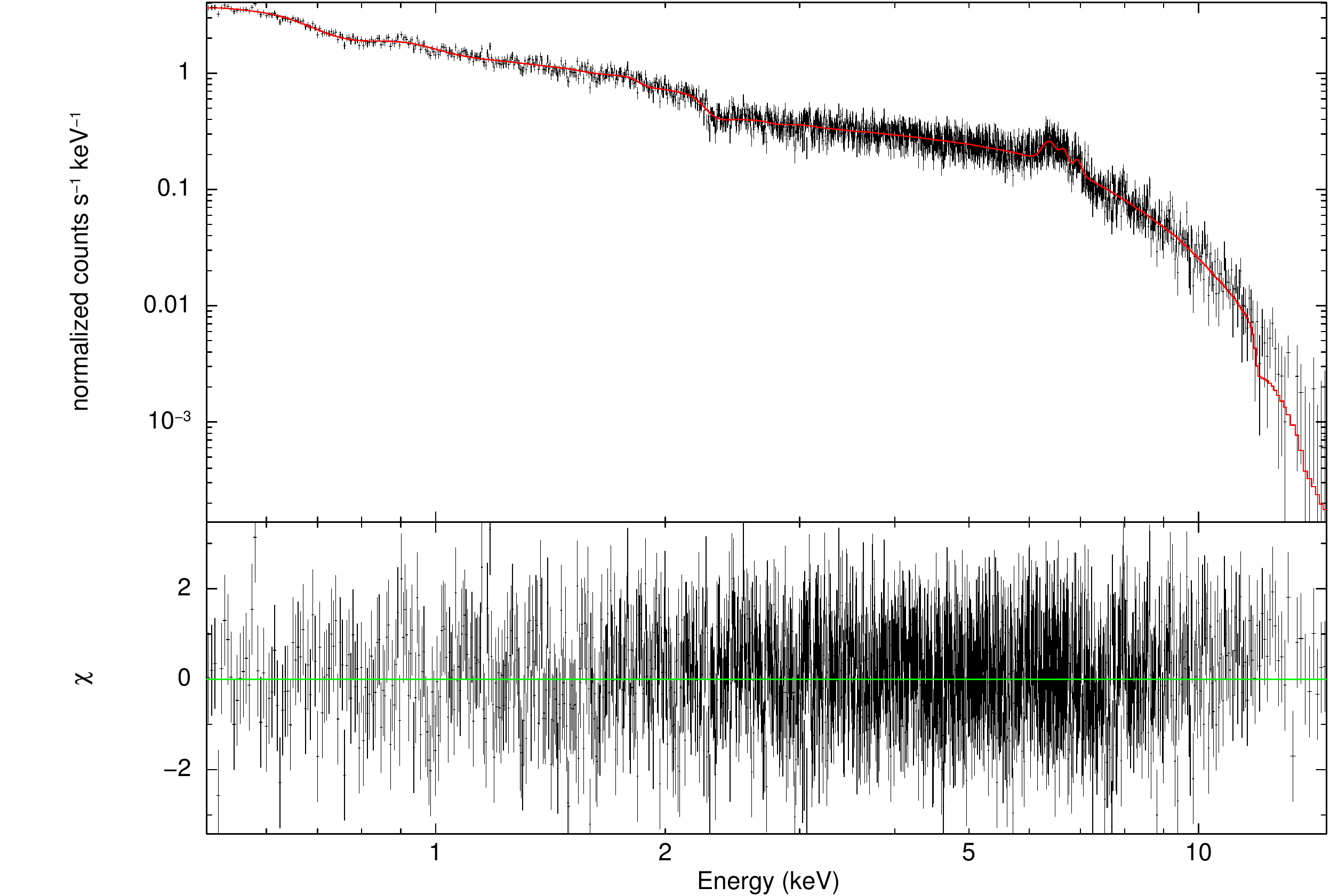}
\includegraphics[width=8.5cm]{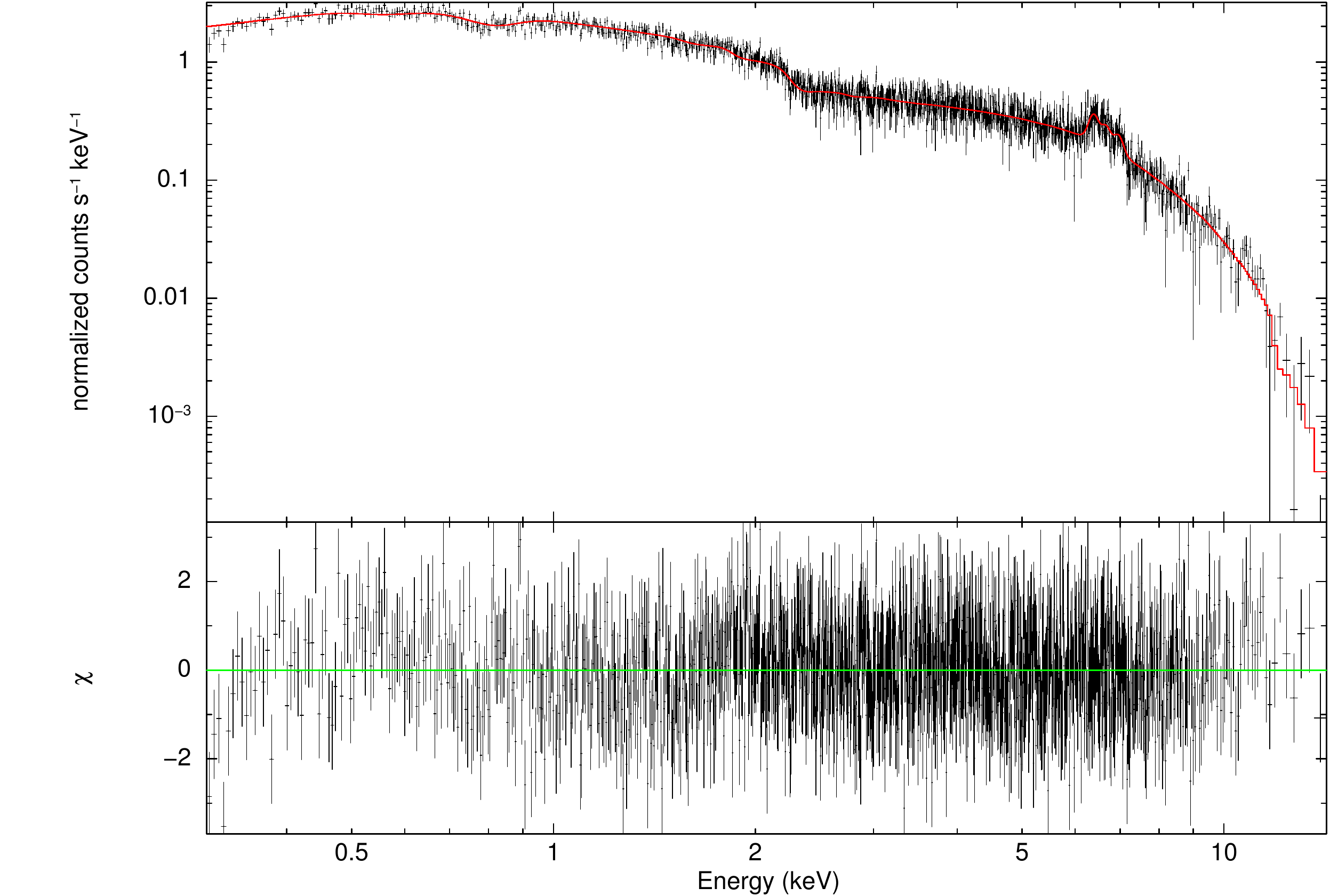}
\caption 
{\xmm\ EPIC pn spectra (\textit{in black}) and the best fit total model (\textit{solid red line}). 
\textit{Left top panel}: spectrum of V2731~Oph in the {1.0 -- 15.0} keV energy range, and the best fit total model \textsc{tbabs*[compTT+gaussian]}.
\textit{Right top panel}: spectrum of RX~J2133.7+5107 in the 0.3 -- 15.0 keV energy range, and the best fit total model \textsc{tbabs*[compTT+(2)gaussian components]}.  
\textit{Left lower panel}: spectrum of PQ~Gem in the 0.5 -- 15.0 keV energy range, and the  best fit total model \textsc{tbabs*[compTT+(5)gaussian components]}.  
\textit{Right lower panel}: spectrum of V2400~Oph in the 0.3 -- 15.0 keV energy range, and the best fit total model \textsc{tbabs*[compTT+(4)Gaussian components]}. See Table \ref{tab:cont}, \ref{tab:Felines}, and \ref{tab:lines}.  
}
\label{fig:spec2XMM}
\end{figure}

\begin{figure}
\centering
\includegraphics[width=8.5cm]{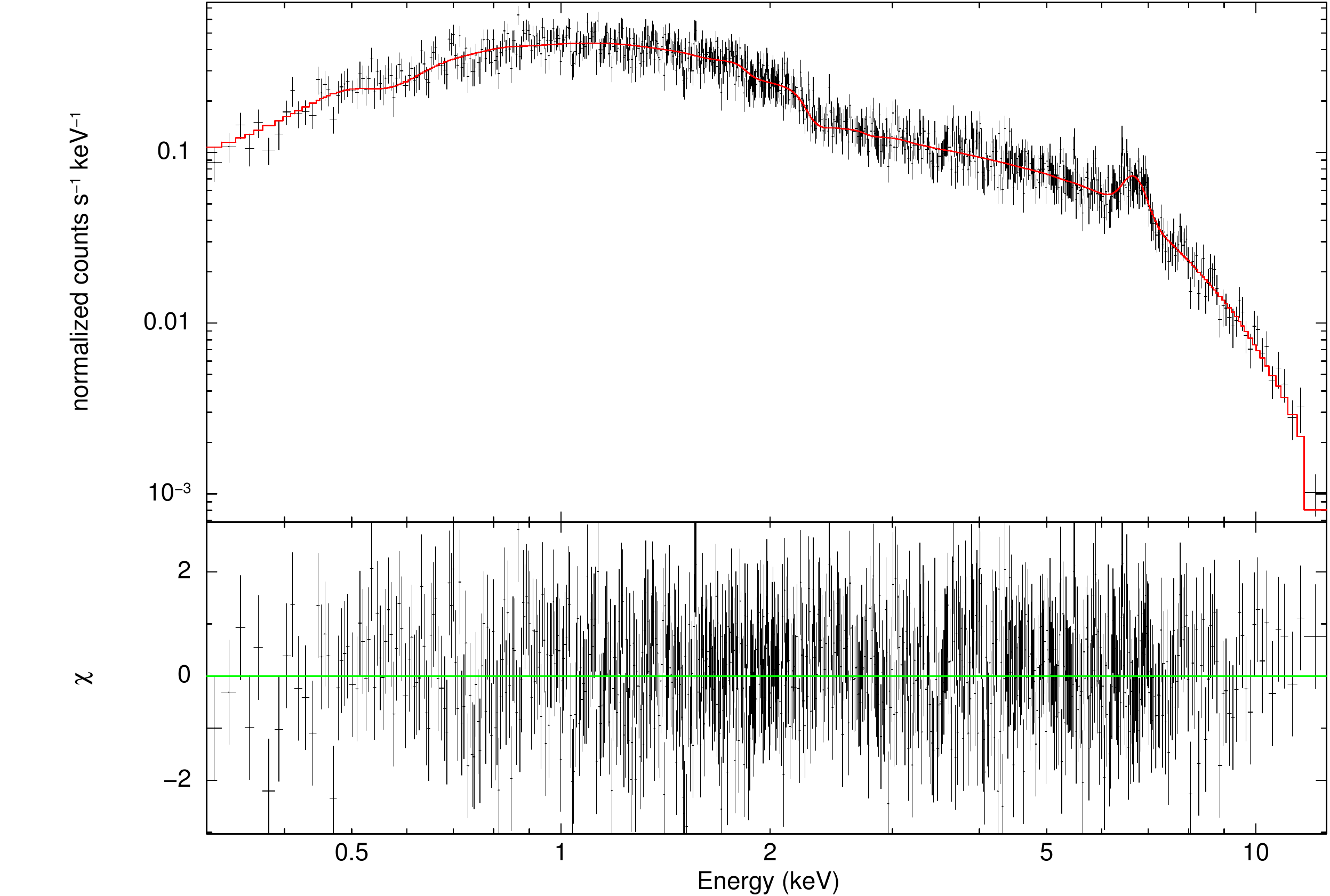}
\includegraphics[width=8.5cm]{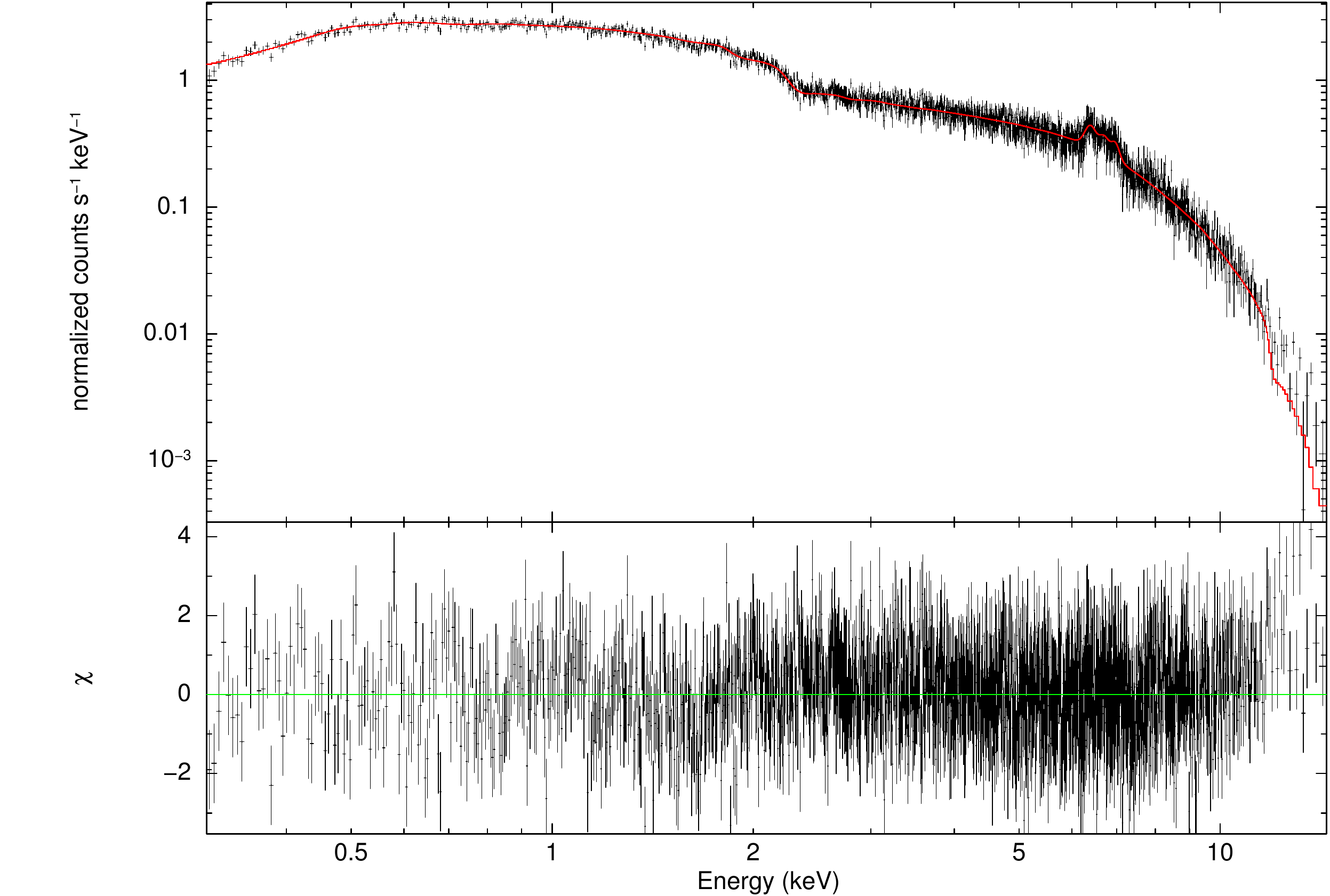}
\includegraphics[width=8.5cm]{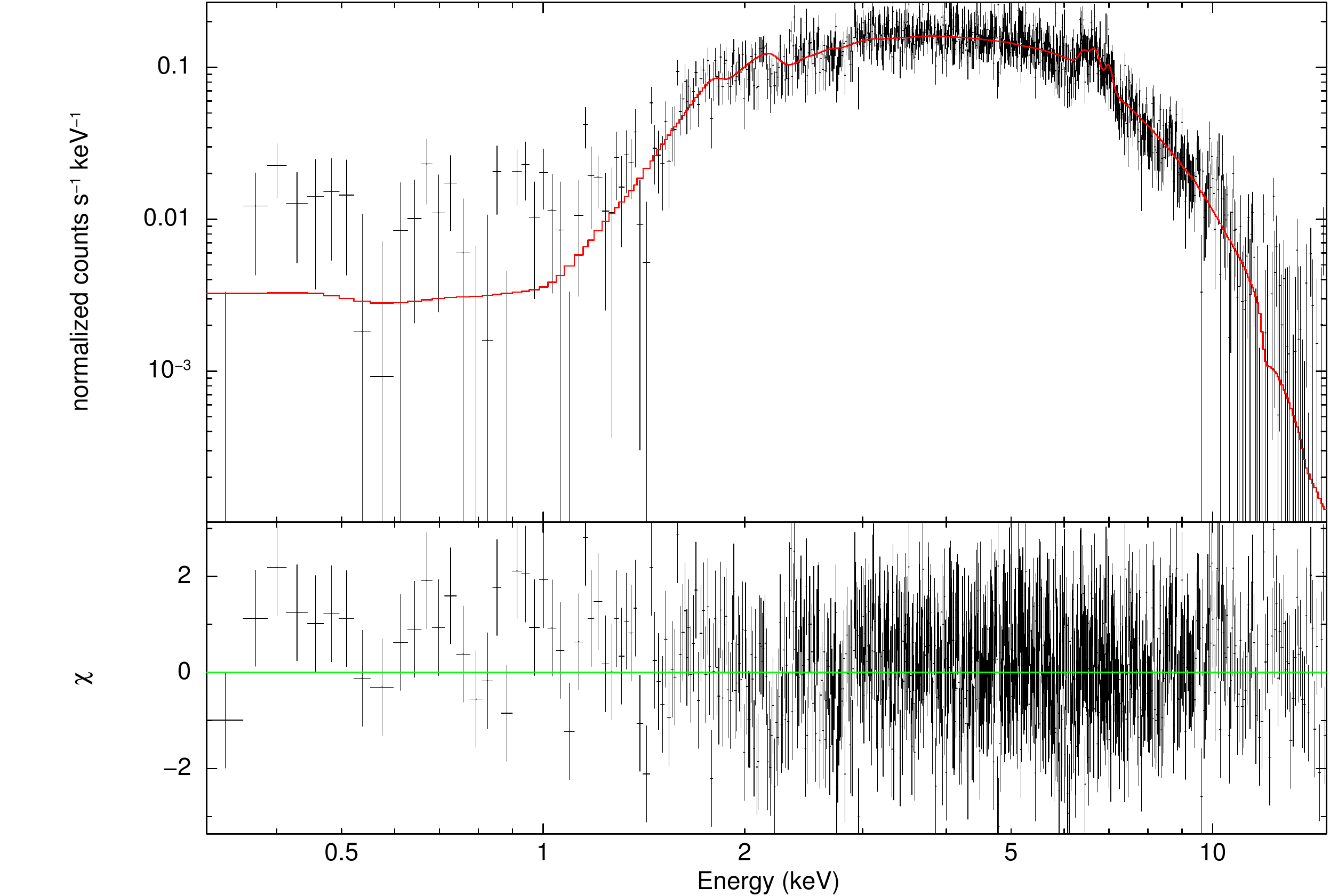}
\includegraphics[width=8.5cm]{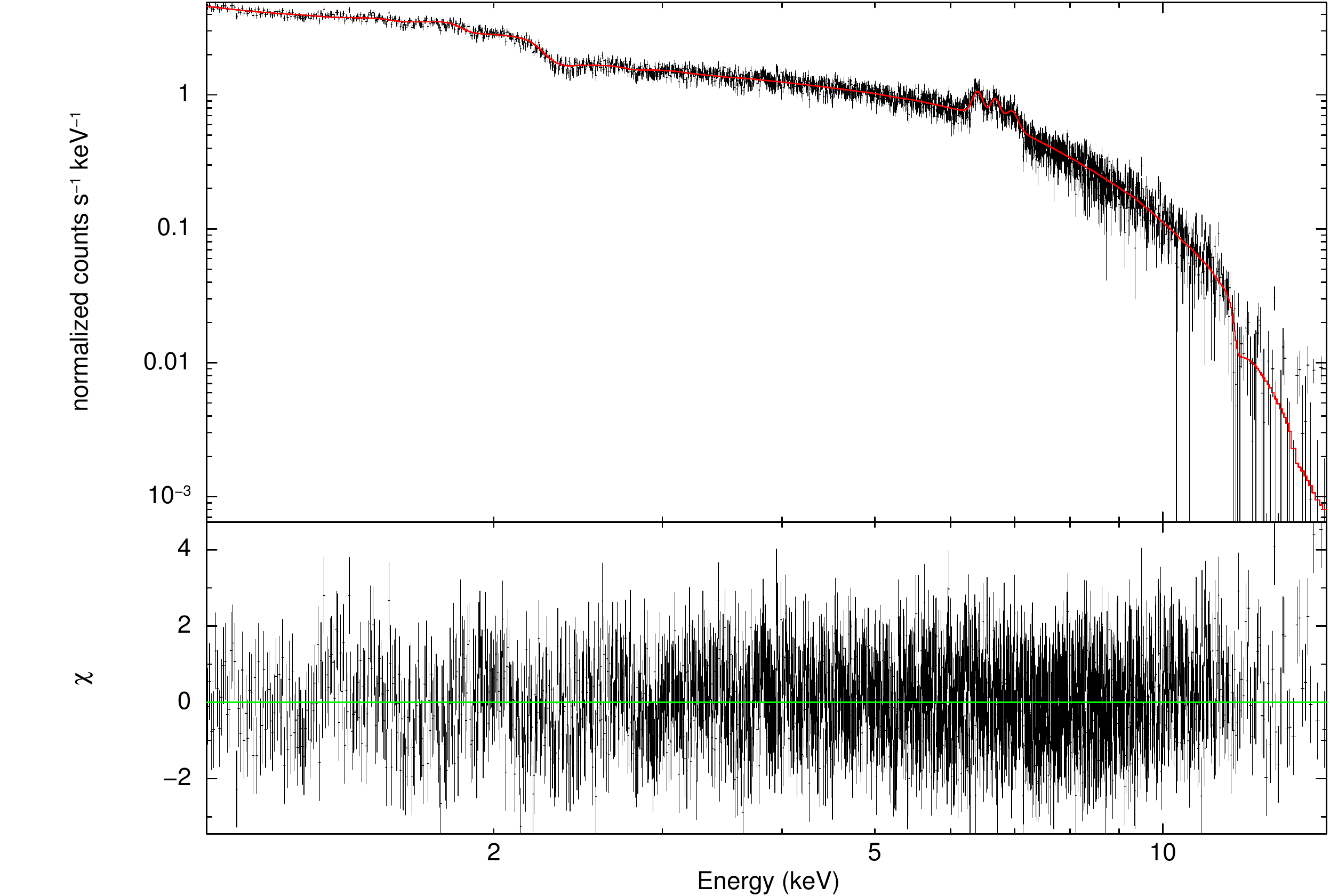}
\caption 
{\xmm\ EPIC pn spectra (\textit{in black}) and the best fit total model (\textit{solid red line}).
\textit{Left top panel}: spectrum of IGR~J00234+6141 in the 0.3 -- 15.0 keV energy range, and the best fit total model \textsc{tbabs*[compTT+gaussian]}.
\textit{Right top panel}: spectrum of IGR~17195-4100 in the 0.3 -- 15.0 keV energy range, and the  best fit total model \textsc{tbabs*[compTT+(4)gaussian components]}. \textit{Left lower panel}: spectrum of XY~Ari in the 0.3 --  15.0 keV energy range, and the best fit total model \textsc{tbabs*[compTT+(3)gaussian components]}. \textit{Right lower panel}: spectrum of V1223~Sgr in the 1.0 --15.0 keV energy range, and the best fit total model \textsc{tbabs*[compTT+(3)gaussian components]}. See Table \ref{tab:cont}, \ref{tab:Felines}, and \ref{tab:lines}.
}  
\label{fig:spec3XMM}
\end{figure}

\begin{figure}
\centering
\includegraphics[width=8.5cm]{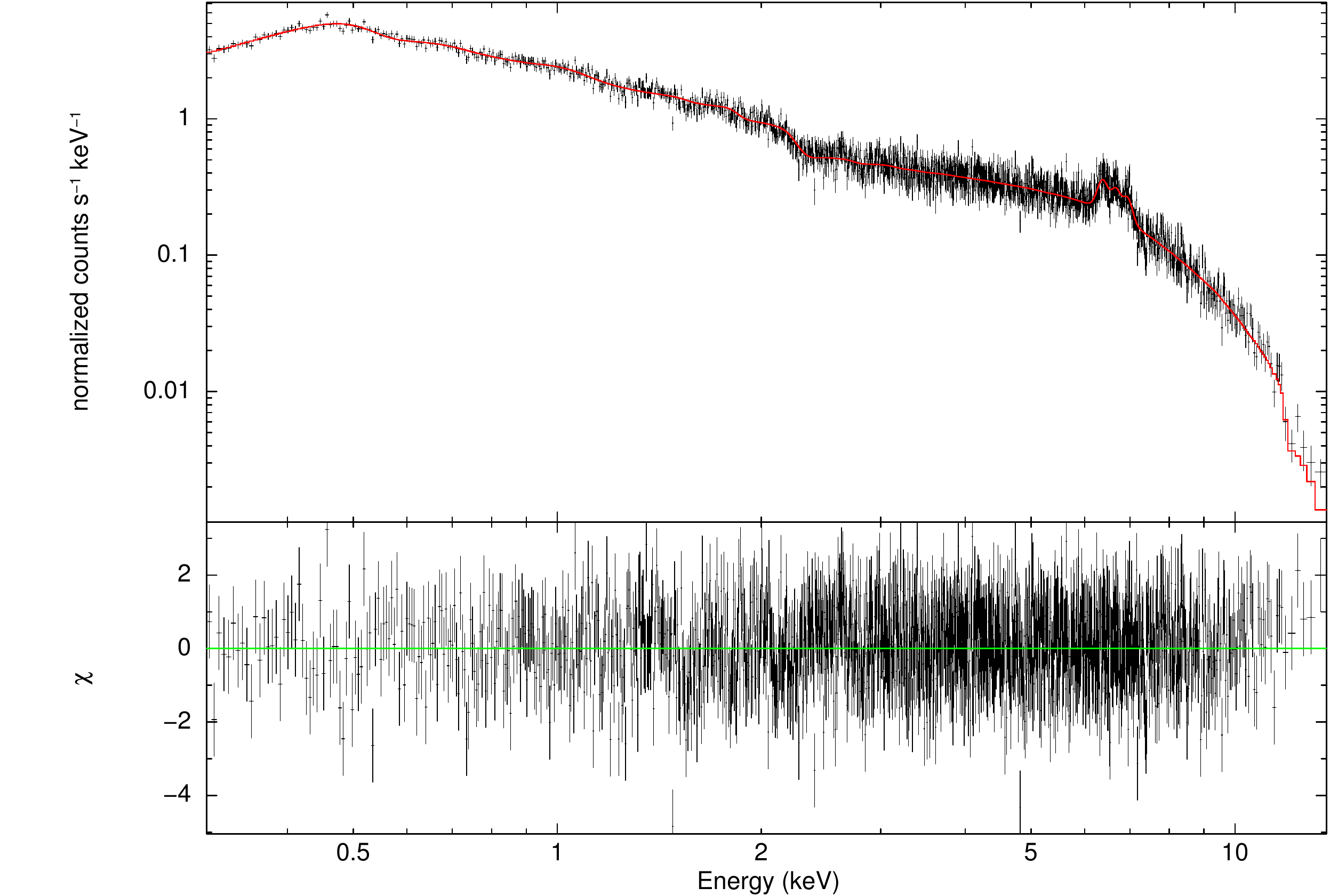}
\includegraphics[width=8.5cm]{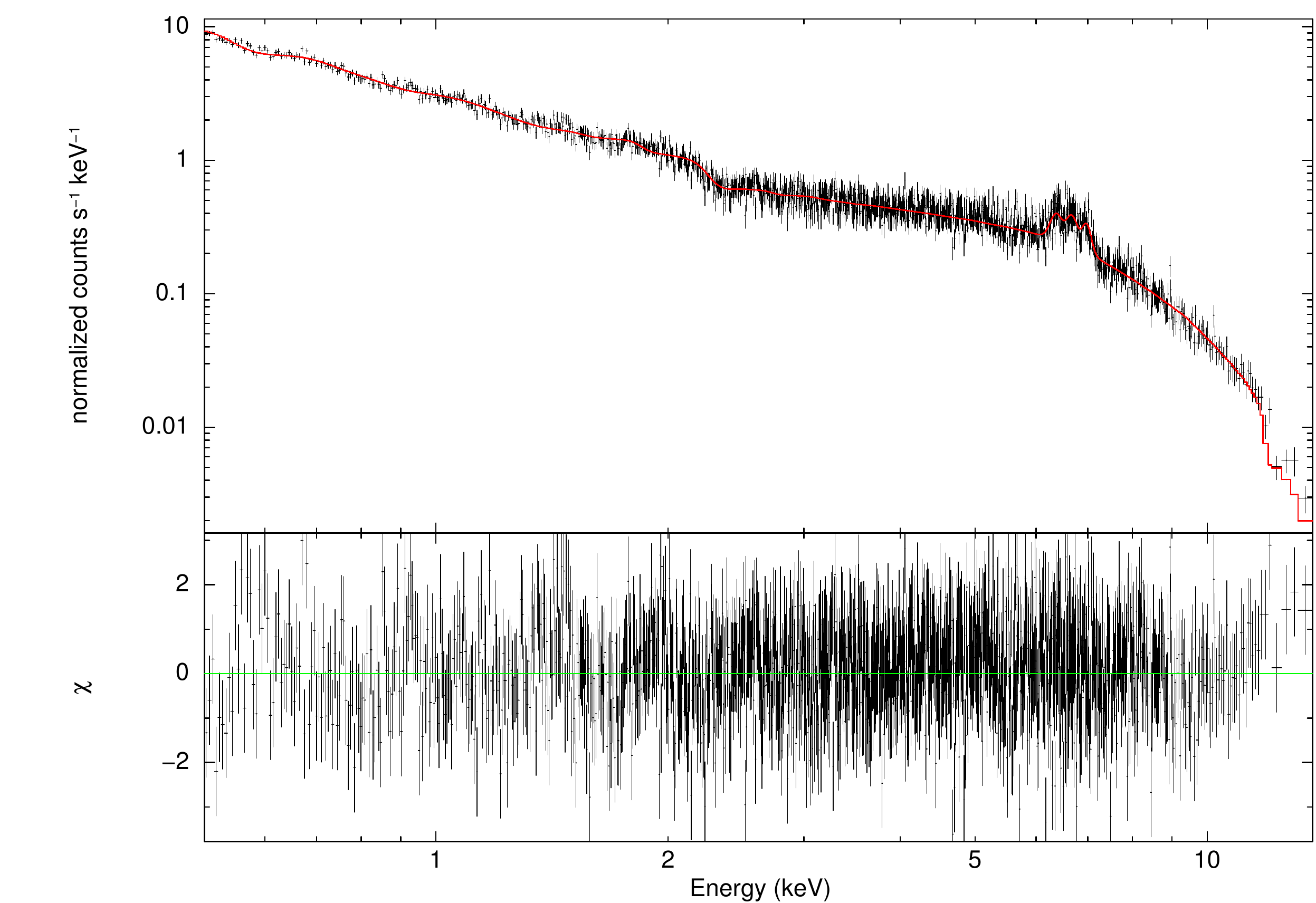}
\caption
{\xmm\ EPIC pn spectra (\textit{in black}) and the best fit total model (\textit{solid red line}). \textit{Left  panel}: spectrum of Obs. ID  0105460301 in the {0.3 --15.0} keV energy range, and the best fit total model \textsc{tbabs*[compTT+(4)gaussian components]}.
\textit{Right  panel}: spectrum of Obs. ID  0105460501  in the {0.5 --15.0} keV energy range, and the best fit total model \textsc{tbabs*[compTT+(4)gaussian components]}. Table \ref{tab:cont}, \ref{tab:Felines}, and \ref{tab:lines}. 
}  
\label{fig:spec4XMM}
\end{figure}

Figure \ref{fig:spec1XMM} shows the best spectral fits in AE~Aqr, V1025 Cen, and HX Hya.  
Figure \ref{fig:spec2XMM} shows the best spectral fits in V2731~Oph, RX~J2133.7+5107, PQ~Gem, and V2400~Oph.   
Figure \ref{fig:spec3XMM} shows the best spectral fits in IGR J00234+6141, IGR 17195-4100, XY Ari and V1223 Sgr.  Finally, Figure \ref{fig:spec4XMM} shows the best spectral fits in NY~Lup Obs. 0105460301 and  0105460501, respectively.

\subsection*{Spectral lines}
For each spectrum, we investigated the presence of each Gaussian component by means of the \textit{simftest} script in \xspec. The probability that data are consistent with the total model without considering each of these components (shown in Table \ref{tab:Felines} and \ref{tab:lines}) is low -- compatible with zero.

In eight out of the thirteen spectra present in our sample we observed all the three emission lines of the $\sim$ 6.4 -- 7.0 keV Fe complex (in EX Hya, PQ~Gem, V2400~Oph, IGR~J17195-4100, XY Ari, V1223 Sgr, and in both observations of NY~Lup; see Table \ref{tab:Felines}). In these observations (a) in general, the Gaussian$_1$ component  is compatible, within errors, with the neutral Fe $\Kalpha$ line at 6.4 keV; only NY~Lup Obs. ID 0105460501 shows a Gaussian$_1$ centroid energy at 1.5 $\sigma$ from the neutral Fe $\Kalpha$ (b) in general, the  centroid energy of the Gaussian$_2$ component is compatible, within errors, with the He-like Fe line at 6.7 keV and/or 6.675 keV; 
EX Hya shows a Gaussian$_2$ with centroid energy at 13 $\sigma$ (due to the small error bar) and 5.3 $\sigma$ from the 6.7 and 6.675 keV lines, respectively, and  V1223 Sgr shows a Gaussian$_2$ line at 2 $\sigma$ from the 6.7 keV line; and (c) only V2400 Oph, IGR~J17195-4100 and XY Ari show a Gaussian$_3$ component with centroid energy compatible, within errors, with the H-like Fe line at 7.0 keV; PQ~Gem, V1223 Sgr and  NY Lup Obs.  ID 0105460301 show Gaussian$_3$ centroid energy at 2 $\sigma$ from this line, while HX Hya and NY Lup Obs. ID 0105460501 show centroid energies at 4 $\sigma$ and 1.5 $\sigma$, respectively. 

In RX~J2133.7+5107 we only observed the Gaussian$_1$ and Gaussian$_3$ components. The centroid energy of the Gaussian$_1$ (appears blue-shifted) at 1.4 $\sigma$ from the neutral Fe line, while the centroid energy of the Gaussian$_3$ (appears red-shifted) at 2.75 $\sigma$ from the H-like Fe line. 
In V1025 Cen, AE~Aqr and IGR~J00234+6141 spectra, we observed only the Gaussian$_2$ component, which appears in general compatible with the He-like Fe line. Only V1025 Cen shows a centroid Gaussian energy apparently blue-shifted at 2 $\sigma$ from the He-like Fe line. However, the peak of this feature is exactly at 6.7 keV. 

Finally, the V2731 spectrum shows only a broad Gaussian$_1$ component, with centroid energy  compatible with neutral Fe line (see Table \ref{tab:Felines}).

AE~Aqr, EX Hya, V1025 Cen, PQ~Gem, V2400, IGR~J17195-4100 and the two observations of NY~Lup show lines other than the Fe lines in the 6.4 -- 7.0 keV energy range (see Table \ref{tab:lines}). These features are present mainly in the soft X-ray spectra (E $\lesssim$ 1.5 -- 2.5 keV). 

In AE~Aqr, we observed a strong and relatively broad excess in the soft X-ray energy band (from $\sim$ 0.7 to 1.1 keV) peaked at $\sim$ 0.9 keV (this peak is compatible with either Ne IX, Ni XIX, Ca XVII, or Fe XIX ion transitions). This feature,  identified as Gaussian$_B$ component in Table \ref{tab:lines}, has a centroid energy at 0.83$_{-0.01}^{+0.01}$ keV, which is compatible with line emission from Fe XVII  
ion at 0.826 keV, or Fe XIX 
at 0.823 keV. This source still show residual trends in the E $< 2.5$ keV energy range (see Figure \ref{fig:spec1XMM}, left top panel),  which may have contributed for a worse spectral fit quality in this source. These residuals are peaked at $\sim$ 1.4 keV, 1.9 keV and 2.4 keV, which could possibly be produced by line emissions from Mg, Si and S ions, respectively. 

V2400~Oph shows an absorption line, identified as Gaussian$_B$ component in Table \ref{tab:lines}. This line shows a centroid energy at 0.808$^{+0.016}_{-0.014}$ keV, which could be absorption of either Ca XVII (at 0.8 keV), Fe XVI (at 0.805 keV), Fe XVII (at 0.802 or 0.812 keV), or Fe XIX (at 0.823 or 0.819 keV) photons. It is important to stress, however, that this centroid energy is close, within errors, to the Fe XVII line energy at 0.826 keV, which is (as previously stated) possibly present in the spectra of AE~Aqr as well.  

As in AE~Aqr, PQ~Gem shows a residual excess peaked at $\sim$ 0.9 keV. The modelling of this feature, identified as Gaussian$_B$ component in Table \ref{tab:lines}, leads to a centroid Gaussian energy at 0.89$^{+0.02}_{-0.02}$ keV. This Gaussian component is compatible with either Ni XIX (at 0.884 or 0.900 keV) or Ca XVIII (at 0.886 keV) ion line emission, and it is very close to Fe XIX (0.918 keV) and Ne IX (at 0.914 keV) ion line emissions. A strong and broad additional trend, peaked at $\sim$ 0.57 keV, is observed in this source. The energy of this peak is compatible with the resonance O VII line at 0.574 keV, and it is close to the K$_{\alpha}$ line of O VII at 0.559 keV. However, the modeling of this trend with a symmetric Gaussian (identified as Gaussian$_A$ in Table \ref{tab:lines}) leads to a centroid energy of 0.52$^{+0.02}_{-0.02}$ keV, compatible with lines from N VII (at 0.500 keV)
and N VI (at 0.522 keV)
 line emissions, and very close to N VI line emission at 0.498 keV.   

IGR~J17195-4100 shows only one line emission with peak and centroid Gaussian  energy at 0.57$^{+0.1}_{-0.1}$ keV (see Gaussian$_A$ component in Table \ref{tab:lines}). In comparison with PQ~Gem, this feature appears rather weak. 

The two NY~Lup spectra show a broad excess peaked at $\sim$ 1.01 -- 1.10 keV (with Gaussian centroid energy at $\sim$ 1 keV in both observations). This line is compatible with Ne X (at 1.022 keV), Fe XVII (at 0.976 and 1.023 keV), Fe XX (at 0.992 and 1.008 keV) and Fe XXI (at 1.008 keV) line emission. It appears stronger in Obs. ID  0105460501 than in Obs.  ID 0105460301 (see Gaussian$_C$ component in Table \ref{tab:lines}). 

The source V1025 Cen shows four line emissions, identified as Gaussian$_B$, Gaussian$_C$, Gaussian$_G$ and  Gaussian$_H$ in Table \ref{tab:lines}. The centroid energy of Gaussian$_B$ is at 1.5 $\sigma$ from the Fe XVII line emission at 0.826 keV. Gaussian$_C$ is compatible within errors with NeX and Fe XXI line emissions at 1.022 and 1.008 keV, respectively. Gaussian$_G$ is compatible within errors with Si XIV line emission at 2.007 keV. Finally, Gaussian$_H$ agrees perfectly with S XV line emission at 2.460 keV.  

Apart from the Fe line complex, the source EX Hya shows ten line emissions, identified by Gaussian$_{C,..,L}$ components in Table \ref{tab:lines}. Gaussian $D$, $E$, $J$, and $L$, agree within errors with Ni XXVI (at 1.348 keV), Mg XII (at 1.473), Ca XIX (at 3.903 keV), and Fe XXVI (at 8.212 keV) line emissions, respectively. Whereas, Gaussian C, F, G, H, I, and K, is at 2.3, 1.7, 2.0, 1.45, 1.3, and 1.05 $\sigma$, respectively, from the Fe XXIII (1.123 keV), 
Si XIII (at 1.865 keV), 
Si XIV (at 2.007 keV), 
S XV (at 2.460 keV), 
S XVI (at 2.623 keV), and 
Ni XXVII (at 7.799 keV) line emissions, respectively.

\section{Discussion} 
\label{sec:disc}
Following Paper I, in this paper we tested if the thermal Comptonization could describe the X-ray continuum of IPs.     
We obtained that the thermal \compTT\ model can, in general,  successfully describe the \xmm\ Epic-pn average continuum of the IPs in our sample. In all 13 observations we obtained \chiqr\ of $\sim$ 1.0 (see Table \ref{tab:cont}). 
However, in five observations we did not obtain a satisfactory fit in the total ~0.3 -- 15 keV \xmm\ Epic-pn energy range. In these cases, our total model is not able to account for the more complex and subtleties of the soft (E $\lesssim$ 1 keV) X-ray spectral shape seen in EX Hya, V2731~Oph, V1223~Sgr, PQ~Gem and NY~Lup Obs. ID 010546050. Our modelling is capable to describe the spectra of V2731~Oph and V1223~Sgr in the 1.0 -- 15 keV energy range, and 
PQ~Gem and NY~Lup in the 0.5 -- 15 keV energy range, respectively -- with best-fitting parameters in agreement with the values found for the other sources. 
The \compTT\ component shows a mean seed photon temperature $<\kts>$ of 0.096 $\pm$ 0.013 keV, 
a mean electron temperature  $<\kte>$ in the Compton cloud of 3.05 $\pm$ 0.16 keV, and 
a mean optical depth $<\tau>$ of 9.5 $\pm$ 0.6. 

\subsection{Soft X-ray component}
We observed an additional soft blackbody (\textsc{bbody}) component in five out of the thirteen spectra in our sample (in EX Hya, RX~J2133.7+5107, V1223~Sgr, and NY~Lup Obs.  ID 010546030\textbf{1} and Obs.  ID 010546050\textbf{1}).  

This \textsc{bbody} component shows a mean temperature $<\ktbb>$ of $0.095 \pm 0.004$ eV, and it is  
in general compatible with values reported in the literature. For example, 
\citet{Anzolin2009} stated the presence of a blackbody with a temperature of $\sim$ 100 eV in  the average (using 0.3 -- 100 keV \xmm, same Obs. ID present in our sample, and \SUZ\ spectra)  spectra of RX~J2133.7+5107. In our analysis this source showed a  blackbody component of $105^{+3}_{-2}$ eV, which is in agreement with the literature. The source NY~Lup is another IP well known for showing a soft blackbody X-ray component  \citep{Potter1997}. Namely, \citet{Harberl2002} using the same \xmm\ Epic-pn observation present in our sample,   found  a blackbody of $\sim$ 84 -- 97 eV. A value of $\sim$ 104 eV \citep{Evans2007,katajainen2010} is reported in the literature as well. In our analyses, we found a \textsc{bbody} component of $95^{+3}_{-4}$ eV and $84^{+2}_{-2}$ eV, respectively, in the two observations of NY Lup analysed (see Table  \ref{tab:cont}, line 15 and 16) --  which are in agreement with the values reported in the literature. 

A soft blackbody component is reported in V2400 Oph as well. For example,  \citet{deMartino2004} using \textit{BeppoSAX} data found  a blackbody component of 103 $\pm$ 10 eV,  \citet{Evans2007} revealed  a temperature of $117^{+33}_{-44}$ eV for this component using the same observation present  in this paper, and  \citet{Josh2019} showed a blackbody component with an average temperature of $\sim$ 98 eV in the \xmm\ and \SUZ\ spectra. However, in our analysis of V2400~Oph  average spectrum a blackbody component was not required by the data. A possible reason for not observing this component could be the different software version used for the data reduction and the different spectral total modelling (framework) used to describe the data. Similarly, in V2731~Oph a blackbody component (of $\sim$ 90 eV) is reported in the literature \citep{Martino2008}, but we did not find this component in our analysis.

As previously stated, a soft blackbody component was not necessary to describe the \xmm\ data of PQ Gem in our final modeling (that is, when Gaussian components were added to the total model to account for emission features in E $\lesssim$ 1 keV). The \textsc{bbody} temperature of $88^{+4}_{-3}$ eV found when considering this component is slightly higher than those pointed out in the literature of $\sim$ 50 -- 60 eV \citep[see][and references therein; who analysed the same data set in the standard multi plasma temperature framework]{Evans2007}. 

\citet{Evans2007}, using the standard spectral modelling, did not state the presence of a blackbody component in the spectrum of V1223 Sgr. In our framework, on the other hand, we observed a weak blackbody with temperature of 100 $\pm$ 10 eV when fitting in the 1.0 -- 15.0 keV energy range. However, when using the total 0.3 -- 15 keV energy band we observed that the temperature of this component is found in the 20 -- 40 eV range, which depends on the modelling of the residual soft (E $\lesssim$ 1 keV) X-ray excesses.  

\citet{Suleimanov2016} stated that the soft excess in the (XIS + HXD) \SUZ\ spectrum of EX Hya can be modelled by either a blackbody or an \textsc{APEC} component with $kT \sim 200$ eV, which is approximately two times of the temperature found by our analysis.  

Therefore, we found  that the presence and the temperature of this soft X-ray component can depend on the modelling of the lines present in the soft energy range, the total spectral model, and (as expected) the spectral energy range analysed. 

\subsection{Continuum spectral shape in IPs}
In general, the shape of a Comptonization spectrum is composed by four parts: a blackbody hump, a power-law, a high energy tail, and a transition region between the parts. 

The energy spectral index $\alpha$ of the spectra in our sample ranges (within errors) from 0.20 to 1.2, with a mean value $<\alpha>$ equal to 0.45 $\pm$ 0.07. 
Namely, we observed a bimodal space for $\alpha$. In  ten observations with the Comptonization parameter $y$ ($y\sim 4\kte \tau^2/\me c^2$) $\gtrsim$ 2 (and $<\tau>$ of 10.4 $\pm 0.4$), we obtained spectral index $<\alpha>$ of $0.34 \pm 0.02$. On the other hand, in the three observations (AE~Aqr, EX~Hya, and V1025 Cen) with $y$ $\lesssim$ 1.0 (and lowest $\tau$ among the sources, with $<\tau>$ equal to $6.5_{-0.4}^{+0.4}$), we obtained the highest $\alpha$, with $<\alpha>$ equal to $0.83 \pm 0.19$. 

\citet{Titarchuk2009}, using 
 the  \textsc wabs*(compTT+Gaussian) model in XSPEC but different spectral energy range and Compton cloud geometry, 
found an acceptable spectral fit (\chiqr = 73.6/77 $\sim$ 1.0) to a \xmm\ spectrum of the IP GK~Per. 
The authors reported higher best-fitting parameters for their thermal \compTT\ component ($\kts$ equal 
to 1.05 $\pm$ 0.57 keV, $\kte$ equal to 5.3 $\pm$ 0.5 keV and $\tau$ equal to 20.9 $\pm$ 1.8; which gives $\alpha$ equal to $0.21^{+0.06}_{-0.04}$). In comparison with our modeling, the different spectral energy 
band and Compton cloud geometry likely contributed to a difference observed in $\kte$ and $\tau$ parameters. 
In particular, if one  uses a spherical geometry then 
 the optical depth  of the Compton cloud increases. For the sources of our sample, we obtained 
$<\tau>$ of $\sim$ 20.
It is important to stress, however, that  different plasma geometry (sphere or plane) does not affect the shape of the resulting spectra.

For optically thick Compton clouds {with $\tau\gg 1$}  
all soft photons illuminating a cloud  are Comptonized. 
Consequently, it can be expected that the power-law portion of the spectrum should  be  relatively small (or even not present at all), and in this case the emergent spectrum is  close to a Wien shape  \citep{Hua1995}.

The photon index $\Gamma$ ($\Gamma = \alpha +1$) of about $1.30$ found in the majority of the sources in our sample may indicate a spectral dichotomy between  nmCVs and the majority of  IPs; the photon indices $\Gamma \sim 1.85$ is reported in nmCVs (Paper I) when broadband spectral analysis is considered. The difference between these spectra lies on the fact that the spectra of the IPs with $\Gamma \sim 1.30$ approaches to a Wien shape \citep{Sunyaev1980}. This lower photon index is likely caused by the increase in the luminosity, which  consequently changes the physical conditions in the Compton cloud (that is, it increases its optical depth).

The sources AE~Aqr, EX~Hya, and V1025 Cen show $<\Gamma>$ compatible, within errors, with that found in nmCVs. In this case the spectral formation is similar to that in nmCVs.

\begin{figure}
\centering
\includegraphics[width=18cm]{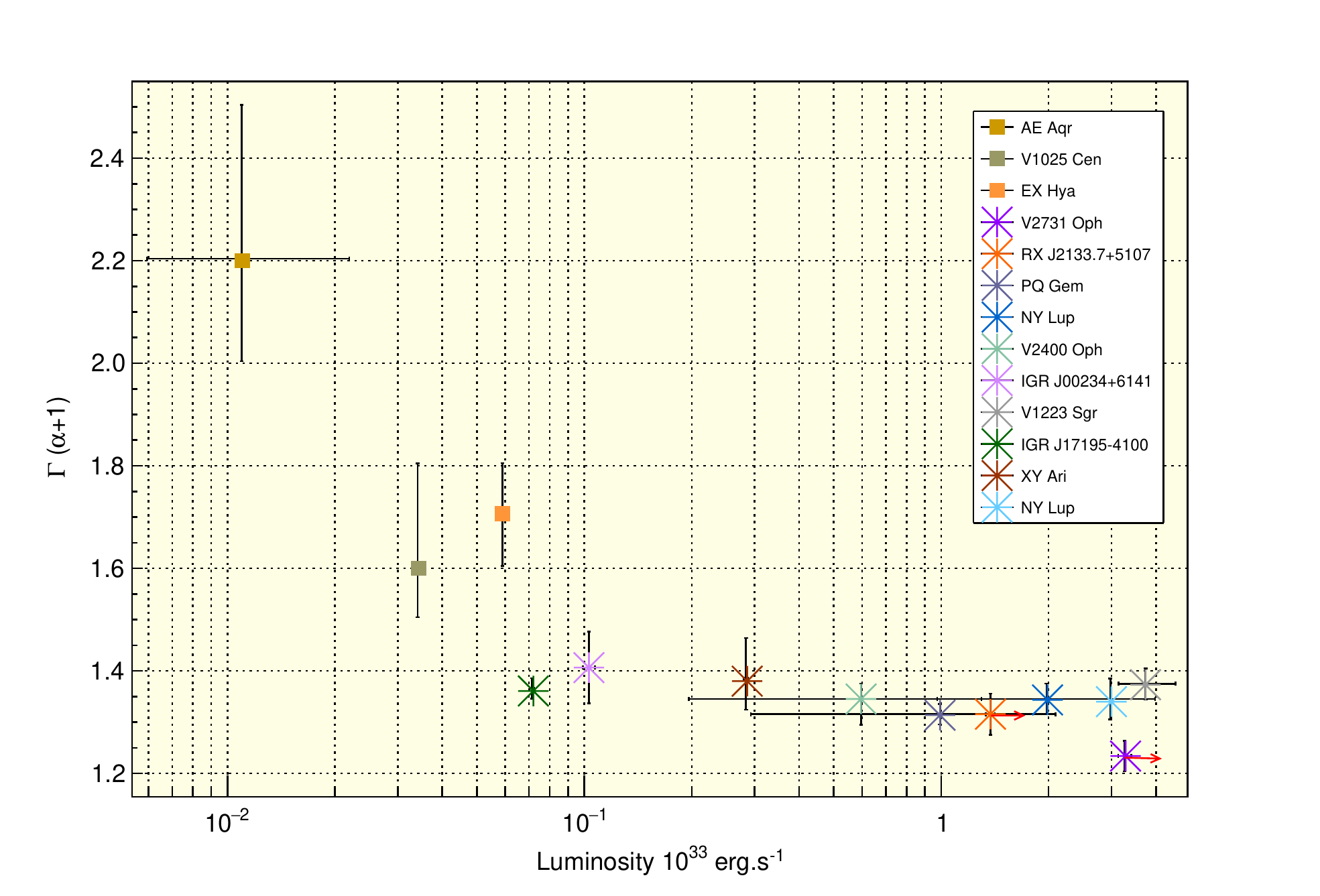}
\caption
{Photon index $\Gamma$ versus luminosity of the unabsorbed Comptonization (\textsc{CompTT}) component in the  
($\sim$ 0.3 to 15 keV) energy band (see Table \ref{tab:cont} and \ref{tab:flux}).  Red arrow indicates lower limit on the luminosity. }
\label{fig:LGamma}
\end{figure}

Table \ref{tab:flux} shows the unabsorbed fluxes of our total modeling, the spectral \textsc{compTT} component (from which luminosity has been computed),   
and the \textsc{bbody} component when present. In Figure \ref{fig:LGamma}, despite (a) the large errors on luminosity estimates due to the large errors on distance estimates and (b) the narrow range of physical Comptonization parameters shared by most of the IPs analysed, one can see a negative correlation between $\Gamma$ and L, where $\Gamma$ decreases with increase in L. 

Namely, $\Gamma \sim$ 1.3 is found for the sources with luminosity of about $\sim $10$^{33}$ erg.\s1, 
whereas  $\Gamma \sim$ 2  is found for the faint IPs with $\sim$ 10$^{31}$ erg.\s1. This bimodal spectral (or photon) index observed in Figure \ref{fig:LGamma}  follows the bimodal X-ray luminosity L observed in IPs, which demonstrated two peaks at L $\sim 10^{31}$ and $\sim 10^{33}$ erg. \s1 \citep{Pretorius2014}.

The source AE~Aqr has magnetic field ($B_{WD}\sim10^{6}$~G) in the lower range of IPs, and shows similarities with nmCV spectra: L $\sim 10^{31}$ erg.\s1, presence of a broad and strong excess in the soft X-ray energy band ($<$ 1 keV) peaked at $\sim$ 0.9 keV  (with centroid Gaussian energy at $0.83_{-0.01}^{+0.01}$ keV), and photon index $\Gamma \sim 2$ -- all characteristics found in nmCV spectra in the thermal Comptonization framework (Paper I). It is important to notice, however, that strong and broad emission lines peaked at $\sim$ 0.9 -- 1.01 keV are also observed in others IPs (see Table \ref{tab:lines}).
Therefore, the high spectral index in the IP AE~Aqr, as well as in EX Hya and V1025 Cen, can be related to close physical conditions in faint IPs and nmCVs for X-ray emission in the Compton cloud and X-ray (re)processing in the (in/out)flow in the system. \citet{Luna2010} stated that EX Hya shows spectral features of both magnetic and nmCVs, when analysing the $\Ch$ spectrum of this source in the standard framework of the post shock region models \cite[see also][]{Mukai2003}. 

Comparing the best-fitting parameters of the  \textsc{compTT} component  found in this present work  with those found in nmCVs through \xmm\ data, we see that 
the temperature of the electrons in the Compton cloud in IPs is about two (or more) times lower ($\kte$ ranges from $\sim$ 5.99 to 8.72 keV in nmCVs) while the optical depth (for plane geometry) is around two (or more) times higher ($\tau$ ranges from 2.65 to 4.73 in nmCVs) (Paper I). 

It is also important to point out that AE Aqr also shares characteristics with young spin-powered pulsars in the 2 -- 20 keV range \citep[see][]{Oruru2012}. AE Aqr is a peculiar nova-like, which hosts the fastest spinning WD ($P_{\text{spin}}=33$~s) \citep{Wynn1997}, which is spinning down at  $\dot{P}_{\text{spin}}=5.64\times 
10^{-14}$~s~s$^{-1}$ rate \citep[and references therein]{vanHeerden2015}. 
\citet{Gotthelf2003} observed that the 2 -- 10 keV  \textit{Chandra} spectra of young rotation-powered pulsars show photon indices such that $0.6 < \Gamma < 2.1$. Interestingly, the nine young rotation-powered pulsars studied by them showed mean photon index $<\Gamma> \sim 1.4$, which is close to the mean value found in this study for IPs.

\subsection{Some examples of previous spectral analyses and comparison with our results}

The standard spectral modelling of IPs is not homogeneous. That is, their spectra have been described by either up to three optically thin plasma temperatures or  cooling flow models --  exactly like in nmCVs. 
 For example, 
the emission component $\leq$10~keV of AE~Aqr has been fitted by multi-temperature models \citep{Terada2008,Oruru2012}.
\citet{Terada2008} using \textit{Suzaku} data, fitted the X-ray spectra of AE~Aqr using multi-thermal components, with temperatures of 
$2.9^{+0.20}_{-0.16}$~keV and $0.53^{+0.13}_{-0.14}$~keV. The excess above the extrapolated hard X-ray emission data was explained by either a third thermal component of $54^{+0.26}_{-0.47}$~keV or a power law with a photon index of 1.12$^{+0.63}_{-0.62}$. 
\citet{Martino2008} described the  0.2 -- 100~keV \xmm\ and \INT\ spectra of V2731~Oph by multi-temperature thermal plasma with  temperature ranging from 0.17 to 60 keV (plus a hot blackbody of $\sim$90~eV, as previously stated). \citet{Anzolin2009} modeled the 0.3 to 100~keV \xmm\ and \INT\ average spectra of RX J2133.7+5107 and IGR~J00234+6141 (their data sample contains the same \xmm\ Epic pn observations presented in this paper). In addition to the optically thick blackbody component of $\sim$ 100 eV (previously mentioned), the authors described the spectra by means of (a) 2 and (b) 3 
different \textsc{mekal} components, and (c) by a multi-temperature \textsc{cemekl}. For the fit considering two \textsc{mekal} components, the authors found a low (kT $\sim$ 0.17 keV) and an intermediate temperature (kT $\sim$ 10 keV). The additional third component shows kT $>$ 27 keV. For the cooling flow \textsc{cemekl} model the authors stated a maximal \textsc{cemekl} temperature in the 40 to 95~keV range \citep[see Table~4 and 5 in][]{Anzolin2009}.

\citet{Harberl2002}, using the same \xmm\ EPIC observation of NY~Lup presented in our sample (Obs. ID 0105460301),  fitted a multi-temperature plasma component to the 0.1--12~keV NY~Lup spectrum, and stated a maximum temperature of $\sim$ 60~keV. \citet{Josh2019}  used two \textsc{MEKAL} components, with $kT_1 \sim$ 7 and $kT_2 \sim$ 26 keV, to explain the 0.3 -- 50 keV \SUZ\ spectra of V2400~Oph; and one \textsc{MEKAL} component with $kT \sim$ 10 keV to describe the \xmm\  spectra. \citet{Xu2016} using \SUZ\ data fitted a single thermal plasma temperature \textsc{apec} model to the 2 -- 10~keV 
continuum of several IPs. 
They stated a plasma temperature of $65.9_{-13.3}^{+18.3}$~keV in RX~J2133.7+5107, $32.6^{+12.7}_{-7.11}$~keV in PQ~Gem, $43.5_{-7.53}^{+11.76}$~keV in NY~Lup, $22.8^{+2.01}_{-1.78}$~keV in V2400~Oph, $26.6^{+4.40}_{-3.47}$~keV in V1223 Sgr, $30.5^{+8.94}_{-6.06}$~keV in IGR J17195-4100, and 
$39.6^{+13.1}_{-8.21}$ and $31.6^{+13.1}_{-8.46}$ keV in  XY~Ari, respectively. 

In comparison with the standard scenario, when only one optically thin thermal plasma temperature is used, the Comptonization model can give temperatures $\sim$ 10 times lower (when they are determined applying  the 0.3 -- 15 keV \xmm\ spectra).

In our spectral modeling we used only a simple photoeletric absorption  component. The presence of dense absorber material partially covering the X-ray source is not required. In the standard framework, however,  
one or more partial covering absorbers components are  commonly necessary in the total modelling for a good description of the data (see section \ref{sec:int}). For example, in V2400~Oph and NY~Lup about $\sim$ 50\% of the X-ray source is stated to be partially covered by a dense absorber \citep[see, \eg][]{Harberl2002,Evans2007,Josh2019}. On the contrary, we see in our modelling that in 6 out of 13 observations the hydrogen column component is attenuated by a $\sim$ 2 -- 20 factor. Only in three  spectra (in EX Hya, XY~Ari and V1223~Sgr; see Table \ref{tab:cont}), the average spectra are absorbed by denser hydrogen column, which indicates to the presence of interviewing dense material likely located somewhere within or close to the system. In 4 observations we obtained  satisfactory fits by setting and freezing the hydrogen column $\NH$ parameter equal to the value given by the HI4PI survey. 

%

\begin{table}
\caption{Unabsorbed flux found for our total modelling, the \textsc{compTT} and \textsc{bbody} components (in the $\sim$ 0.3 to 15 keV; see Table \ref{tab:cont},\ref{tab:Felines},\ref{tab:lines}).}
\label{tab:flux}
\begin{center}
\small
\begin{tabular}{llllll}
\toprule
Source & Flux (total) &Flux (compTT) &Flux (blackbody) & Distance & Refs\\
	   &$10^{-11}$ erg.\cm2.\s1 &$10^{-11}$ erg.\cm2.\s1 &$10^{-11}$ erg.\s1.\cm2 & pc & \\
\midrule
AE Aqr &1.088$^{+0.021}_{-0.011}$ &$0.859^{+0.011}_{-0.009}$ & &$102^{+42}_{-23}$& 1\\
EX Hya &13.56$^{+0.03}_{-0.03}$ &11.85$^{+0.04}_{-0.04}$ &0.707$^{+0.011}_{-0.011}$ &64.5$^{+1.2}_{-1.2}$ & 2\\
V1025 Cen &2.481$^{+0.022}_{-0.022}$ &2.374$^{+0.026}_{-0.026}$ & &110 & 2\\
V2731 Oph &2.85$^{+0.11}_{-0.10}$ &$2.75^{+0.11}_{-0.11}$ & & $>1000$ & 2\\
RX J2133.7+5107 &4.25$^{+0.08}_{-0.08}$ &$3.22^{+0.08}_{-0.10}$ &$0.94^{+0.03}_{-0.04}$ & $>$600 & 3\\
PQ Gem &3.27$^{+0.06}_{-0.06}$ &$3.05^{+0.02}_{-0.02}$ & &510$^{+280}_{-180}$ & 2\\
NY Lup$^{a}$ &4.62$^{+0.05}_{-0.05}$ &$3.89^{+0.08}_{-0.08}$ &$0.57^{+0.008}_{-0.008}$ & $690^{+150}_{-150}$ & 4\\
NY Lup$^{b}$ &9.33$^{+0.05}_{-0.05}$ &$5.20^{+0.04}_{-0.04}$ &$3.85^{+0.05}_{-0.05}$ & &\\
V2400 Oph &6.48$^{+0.13}_{-0.12}$ &$6.63^{+0.05}_{-0.05}$ & &$280^{+150}_{-100}$ & 2\\
IGR J00234+6141 &$0.99^{+0.04}_{-0.04}$ &$0.96^{+0.04}_{-0.04}$ & &$300$ &5 \\
IGR J17195-4100 &$5.16^{+0.05}_{-0.05}$ &$5.01^{+0.07}_{-0.07}$ & &$110$ & 6\\
XY Ari &$3.34^{+0.04}_{-0.04}$ &$3.28^{+0.05}_{-0.05}$ & & & 7\\
V1223 Sgr &$11.9^{+1.1}_{-1.2}$ &$11.49^{+0.04}_{-0.04}$ &$0.162^{+0.005}_{-0.005}$ &$527^{+54}_{-43}$ & 2,5,8,9\\
\bottomrule 
\end{tabular}
\end{center}
References: 1. \citet{fried1997}; 2. \citet[][and references therein]{Pretorius2014}; 3. \citet{Bonnet2006}; 4. \citet{deMartino2006b}; 5. \citet[][and references therein]{Barlow2006}; 6. \citet{Masetti2006}; 7. \citet{Littlefair2001}; 8. \citet{Beuermann2004}; 9. \citet{Nwaffiah2014}.\\
$^a$ Obs. ID 0105460301.\\  
$^b$ Obs. ID 0105460501.
\end{table}

\subsection{The Comptonization scenario in IPs and future perspectives}
\begin{figure}
\centering
\includegraphics[width=8.5cm]{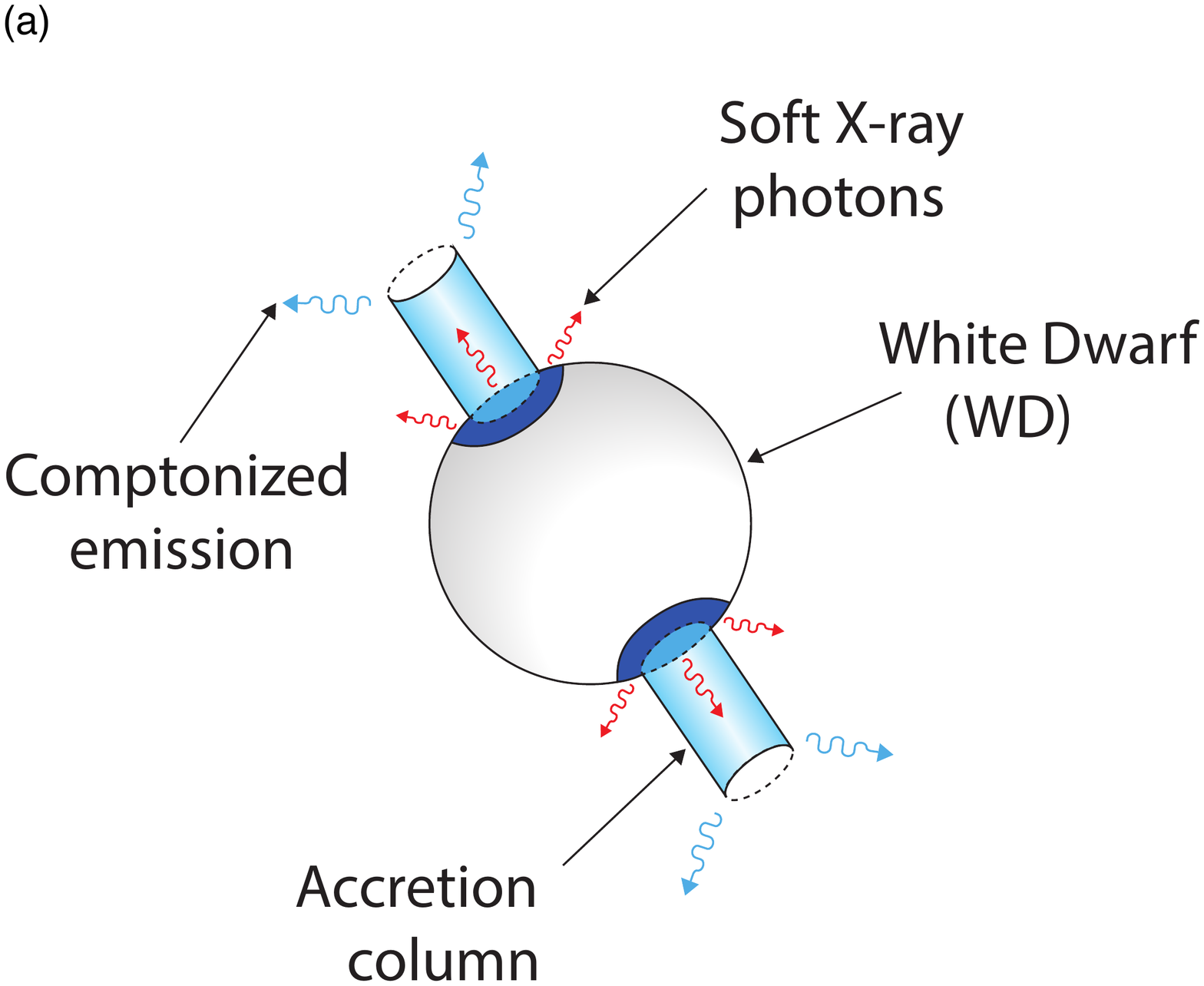}
\includegraphics[width=8.5cm]{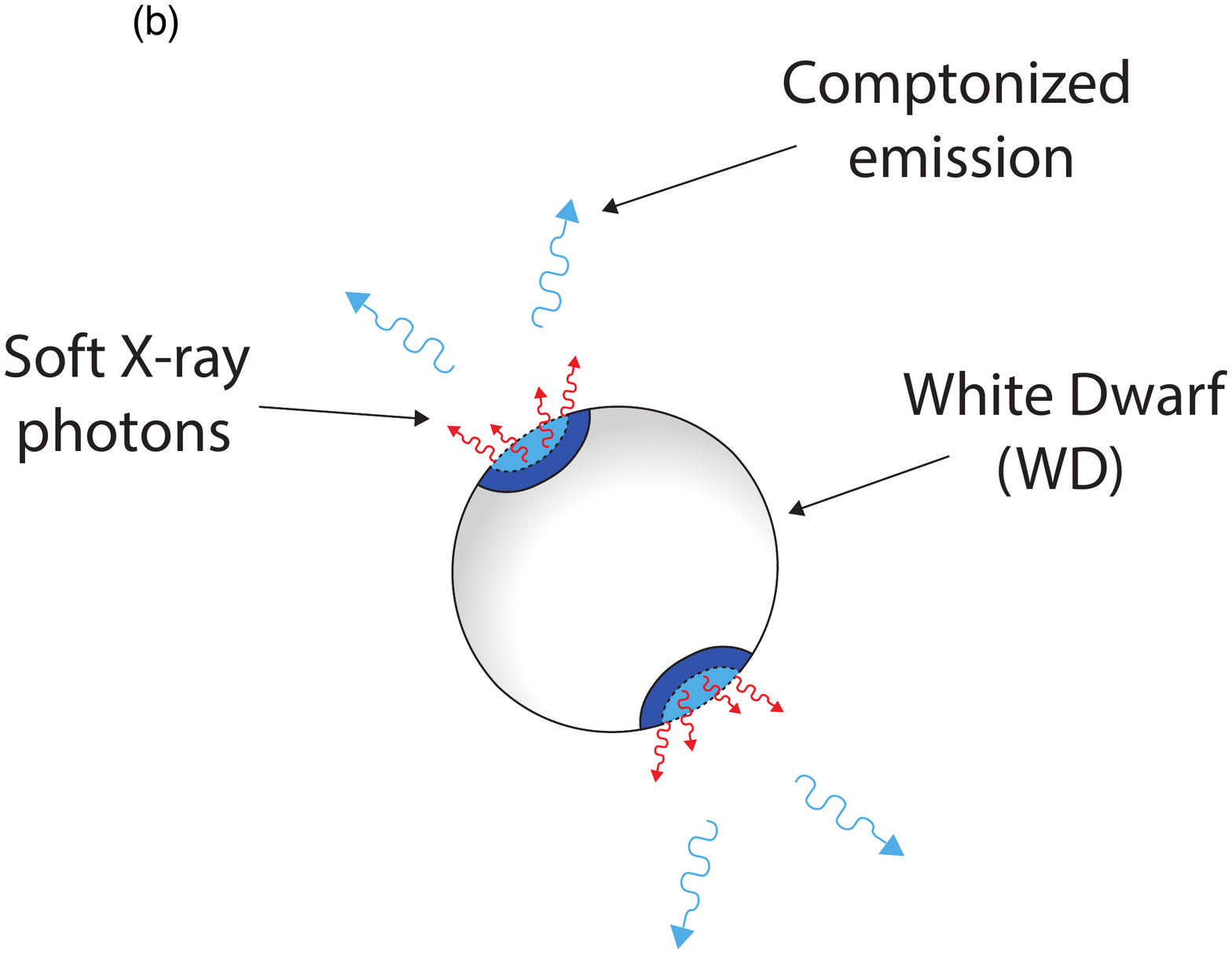}
\caption
{(a) The general scenario of Comptonization in IPs:  UV/soft X-ray photons from the star surface (\textit{red arrows}) are Comptonized by electrons in the optically thick plasma present in the accretion columns. (b) The accretion column is destroyed if the local luminosity on the top of the polar caps is greater than the Eddington. In this case, the UV/soft X-ray photons coming from the star surface are Comptonized by the optically thick plasma (Compton cloud) around to the seed photon emission sites.}
\label{fig:cttframe}
\end{figure}

 In our interpretation, the UV/soft X-ray seed photons characterized by their color temperature $\kts$ in the $\sim$ 15 -- 210 eV energy range are Comptonized by hot electrons of $<\kte> 3.05 \pm 0.16$ keV in  the Compton plasma around the WD, {which is presumably the accretion columns themselves.} 
 The seed photons are likely coming from the WD surface (see Figure \ref{fig:cttframe}a). {In the case of high luminosity ($\Gamma \sim 1.3$), a possible scenario is that the accretion columns are destroyed by high radiation pressure caused by high specific accretion rate (accretion rate per unit area) on the polar cap, and as a result emission escapes from a quasi-spherical region  formed on the star surface (see Figure \ref{fig:cttframe}b).}

 The additional soft blackbody 
 component characterized by $<\ktbb>$ of 95 $\pm$ 4 eV, as previously mentioned, is a well known component necessary to account for the soft X-ray emission of some IPs. This component likely  corresponds to photons coming from the heated region on the WD surface or photosphere -- from the region near to the magnetic pole caps, where the accreting material is dumped, or somewhere else if dense blobs of matter penetrate into the WD photosphere. 
  Part of these soft photons is subject to be Comptonized {by the accretion column (the optically thick plasma located nearby to  the soft seed photons site)}.

Interestingly, a good spectral fit was obtained for the PQ Gem and V2400 Oph. These both systems 
share some characteristics with the polar subclass of mCVs. Namely, PQ~Gem is an unusual mCV because 
it shows a strong soft X-ray component, it is polarized in the optical/infrared (IR) wavebands \citep{Potter1997}, and it has an estimated magnetic field strength $B \sim8-21\times 10^{6}$~G \citep{Vaeth1996,Piirola1993} --- which are characteristics of polars. On the other hand, it shows a spin period of 833.4~s \citep{Mason1997} and an orbital period of 5.19~h \citep{Hellier1994} --- which are typical of IPs.  V2400~Oph is the most polarized IP, leading to the highest magnetic field estimate for any IP ($B\sim9 - 27 \times 10^{6}$~G), overlapping with the range for low-field polars \citep{Buckley1995,Vaeth1997}. In addition, \citet{Buckley1995} showed a strong evidence for a disk-less accretion mechanism in this source. They found that the polarized light varies at 927~s, which was interpreted as the WD spin period. 
The X-rays are pulsed at 1003~s, corresponding to the beat period between the orbital (3.42~h) and spin cycles. The source does not show any X-ray modulations at the orbital or spin cycles, and the detection of the beat period is expected to be observed in disk-less IPs, showing the signature of pole-flipping accretion  
(see \citet{Josh2019} and references therein for a discussion on the nature of the accretion mechanism in this source). 

{It is worthwhile to emphasize  that  the presence of an accretion disk is not essential when one uses the Comptonization model.} 
Therefore, we conclude that the good spectral fits of  the PQ Gem and IP V2400 Oph spectra is an evidence that the thermal Comptonization may describe the spectral continuum of the polar subclass of mCVs as well. 

It is important to stress however that this study is energy band limited, considering only $\sim$ 0.3 -- 15 keV \xmm\ spectra. A future study 
including broader spectral band (E $>$ 15 keV) may better constrain the range of physical parameters in the Comptonization framework and consequently the spectral indices. Therefore, the analysis of a bigger sample 
of IP spectra --  in different states and WD spin phases --  considering data from broadband X-ray missions is highly suitable for a complete characterization and understanding of IPs (and CVs in general) in the Comptonization framework. 
 
\section{Summary and Conclusions}
\label{sec:conc}
We found  that the thermal \compTT\  model can in general successfully fit the continuum of all IPs (13 \xmm\ Epic-pn observations) in our sample. In this scenario the continuum is produced by seed photons with a mean color temperature $<\kts>$ of 0.096 $\pm$ 0.013 keV being Comptonized by electrons with a mean plasma temperature $<\kte>$ 3.05 $\pm$ 0.16 keV located in an optically thick plasma around the central source. {In our interpretation,  the UV/soft X-ray seed photons are coming from 
the WD surface (most likely from the region surrounding the magnetic polar caps) and are being  Comptonized by hot electrons of  (a) the accretion columns or (b) the Compton cloud around the magnetic pole caps of the star. This last case may occurs when accretion columns are not present, namely when $\Gamma \sim 1.3$ and the source has high specific accretion rate.}

The spectra of EX Hya, NY~Lup, V1223~Sgr and RX~J2133.7+5107 show an additional soft X-ray \textsc{bbody} component, with mean temperature $<kT_{bb}>$ of 95 $\pm$ 4 eV, necessary to account for the soft X-ray emission in their spectra. We observed that the presence of this component can strongly depend on both the spectral modelling of the lines in the $\sim$ 0.5 -- 1.2 keV soft energy band and the total spectral modeling. This emission likely comes from the star surface, from the heated region surrounding the star magnetic pole caps. {Thus, a portion of these photons may be Comptonized by the electrons in the accretion columns or Compton cloud located nearby, surrounding, and on top to the soft seed photon  emission regions.}

Furthermore, we observed a dichotomy in the spectral shape of IPs, which is driven by the source luminosity and optical depth of the Compton plasma -- since $<\kts>$ and $<\kte>$ does not change between these two groups of spectra.  IPs of high  luminosity ($L \sim 10^{32-33}$ erg.s$^{-1}$) and $\tau$ ($<\tau>$ of $10.4 \pm 4$) show relatively small $\Gamma$ ($<\Gamma>=1.34 \pm 0.02$; {which indicates the saturated Comptonization in the Compton cloud}). Whereas IPs of lower luminosity ($L \sim 10^{31}$ erg.s$^{-1}$) and $\tau$ ($<\tau>$ of $6.5 \pm 0.4$) show higher $\Gamma$. In this case, $<\Gamma>$ is equal to $1.83 \pm 0.19$, which is exactly what is observed in nmCVs. This may indicate closer (intermediate) physical conditions between faint IPs and nmCVs for X-ray emission in the Comptonization framework.

The good spectral fits obtained for the PQ~Gem and V2400 Oph sources indicate that thermal Comptonization may also describe the spectral continuum of the polar sub-class of mCVs.

\section*{Acknowledgments}
{
T.M.\ acknowledges the financial support given by the Erasmus Mundus Joint Doctorate Program by Grants Number 2013-1471 from the agency 
EACEA of the European Commission, and the INAF/OAS Bologna. 
T.M.\ acknowledges the support given by the NSFC (U1838103 and 11622326) and the National Program on 
Key Research and Development Project (Grants No. 2016YFA0400803). T.M.\ also thanks the High Energy Astrophysics group of the Physics 
Dept.\ of the University of Ferrara, INAF/OAS Bologna and Wuhan University, and Arti Joshi 
for the fruitful discussion on CVs. 
L.T. also acknowledges a support of this work by the astrophysical group in Lebedev Physical Institute of the Russian Academy of Science.
}

\bibliographystyle{apj}
\bibliography{biblio.bib}
\end{document}